# Revealing the Superior Electrocatalytic Performance of 2D Monolayer WSe$_2$ Transition Metal Dichalcogenide for Efficient H$_2$ Evolution Reaction

## Srimanta Pakhira[1,2,3*] and Vikash Kumar[1]

[1] Department of Physics, Indian Institute of Technology Indore, Khandwa Road, Simrol, Indore-453552, MP, India.

[2] Department of Metallurgical and Materials Science (MEMS), Indian Institute of Technology Indore, Khandwa Road, Simrol, Indore-453552, MP, India.

[3] Centre for Advanced Electronics (CAE), Indian Institute of Technology Indore, Khandwa Road, Simrol, Indore-453552, MP, India.

*Corresponding author: spakhira@iiti.ac.in (or) spakhirafsu@gmail.com

**ABSTRACT:** H$_2$ evolution reaction (HER) requires an electrocatalyst to reduce the reaction barriers for the efficient production of H$_2$. Platinum-group metal (PGM) elements such as Pt, Pd, etc. and their derivatives show excellent electrocatalytic activity for HER. The high cost and lack of availability of PGM elements bring constraints over their wide commercial applications, so discovering noble metal-free electrocatalysts with lower possible reaction barriers is paramount important. Two-Dimensional Transition Metal Dichalcogenides (2D TMDs) have emerged as a pinnacle group of materials for many potential applications, including HER. In this work, we have computationally designed a pristine 2D monolayer tungsten diselenide (WSe$_2$) TMD using the first principle-based hybrid Density Functional Theory (DFT) to investigate its structural, electronic properties and the electrocatalytic performance for HER. The possible Volmer-Heyrovsky and Volmer-Tafel reaction mechanisms for HER at the W-edge of the active site of WSe$_2$ were studied by using a non-periodic finite molecular cluster model W$_{10}$Se$_{21}$. Our study shows that the pristine 2D monolayer WSe$_2$ follows either the Volmer-Heyrovsky or the Volmer-Tafel reaction mechanisms with a single-digit low reaction barrier about 6.11, 8.41 and 6.61 kcal/mol during the solvent phase calculations of H$^\bullet$-migration, Heyrovsky and Tafel transition (TS) states, respectively. The lower reaction barriers, high turnover frequency (TOF) ~ 4.24 x 10$^6$ sec$^{-1}$ and 8.86 x 10$^7$ sec$^{-1}$ during the Heyrovsky and Tafel reaction steps and the low Tafel slope 29.58

mV.dec$^{-1}$ confirm that the pristine 2D monolayer WSe$_2$ might be a promising alternative to PGM based electrocatalyst.

**GRAPHICAL ABSTRACT:**

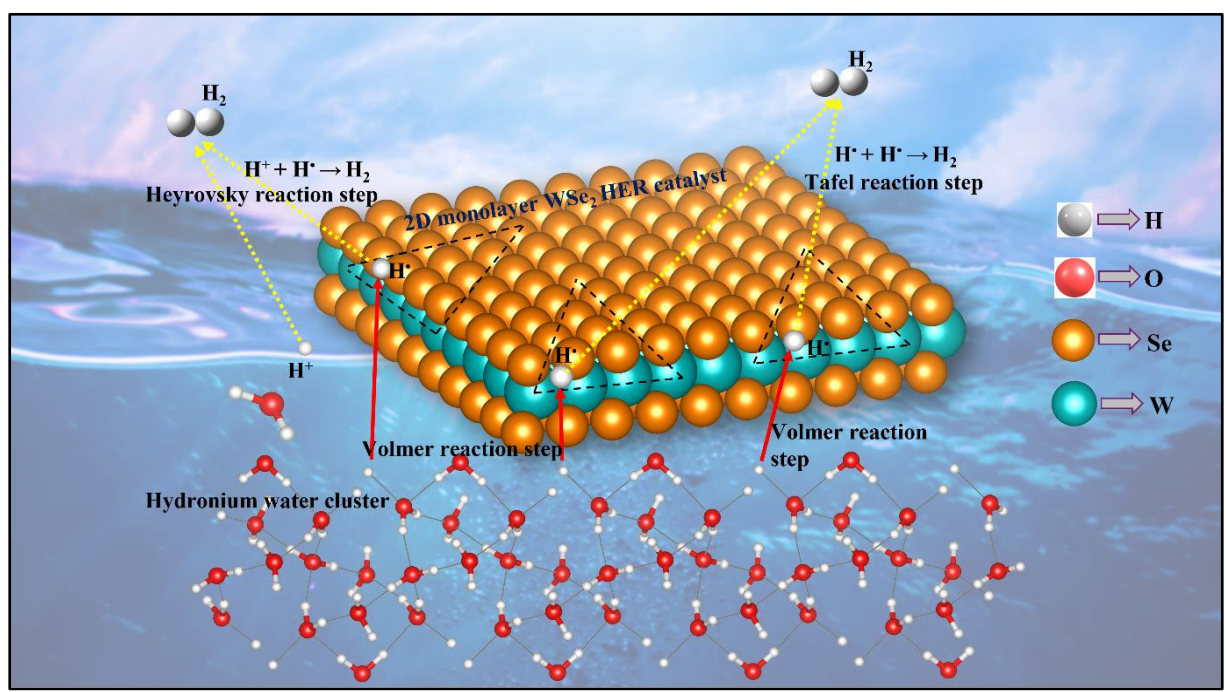

## 1. INTRODUCTION

From last many centuries, the fossil fuel is used as a common energy source to satisfy the requirements of energy supply. However, its consumption is responsible for the emission of many hazardous environmental pollutants and greenhouse gases, resulting in significant climate change. It has an adverse effect on human health as the emissions are very harmful for our society. Therefore, there is an increasing demand for the exploitation of inexpensive, earth-abundant, clean and renewable energy sources that can replace the overutilization of carbon-based fossil fuels and meet the global energy crisis deprived of the emission of any environmental and health pollutants.[1,2] Hydrogen has significant energy content, carries high energy density in its molecular form (H$_2$) and is present enormously in the form of water and other organic and inorganic compounds. Thus, it is considered as a promising and efficient green energy carrier.[3–5] In today's world, most commercially available $H_2$ fuel is manufactured by the steam reforming process. In this process, $H_2$ is produced along with carbon monoxide (CO) and carbon dioxide ($CO_2$) emissions like poisonous gases. The demerit of this process is that it has low energy efficiency for the conversion process and emits carbon-based unsafe by-products.[6–8] An alternative way to produce clean and renewable $H_2$ is through the electrolysis of water with the passage of electricity through it.[7,9]

Overall electrolysis of water completes with two half-cell reactions: hydrogen evolution reaction (HER) at the cathode and oxygen evolution reaction (OER) at the anode.

$$2H_2O_{(l)} \rightarrow 2H_{2(g)} + O_{2(g)}$$

HER is the half key reaction which includes the following steps;[10,11]

Volmer reaction:

$$H^+ + M + e^- \rightarrow MH_{ads} \quad \text{(In acidic medium)}$$

$$H_3O^+ + M + e^- \rightarrow MH_{ads} + H_2O \quad \text{(In alkaline medium)}$$

Heyrovsky reaction:

$$MH_{ads} + H^+ + e^- \rightarrow M + H_2 \quad \text{(In acidic medium)}$$

$$MH_{ads} + H_3O^+ + e^- \rightarrow M + H_2O + H_2 \quad \text{(In alkaline medium)}$$

Tafel reaction:

$$2MH_{ads} \rightarrow 2M + H_2$$

Where M stands for the active site of the catalyst and $MH_{ads}$ stands for the adsorbed hydrogen ($H_{ads}$) at the catalyst's active site. The complete evolution process of molecular hydrogen ($H_2$) may follow either Volmer-Heyrovsky or Volmer-Tafel reaction pathways. An overpotential is required for the adsorption of the proton ($H^+$) at the active site of the electrocatalyst and reduce it into $H_2$ during the above-mentioned reaction pathways.[12] Reduction, in this overpotential, is the crucial part and the same can be obtained with the use of an excellent electrocatalyst for the efficient production of $H_2$. Due to nearly zero overpotential, high exchange current density and small Tafel slope of the platinum-group elements (PGE) i.e. noble metal-based electrocatalysts (such as platinum (Pt), Pd, etc.) in the acidic electrolyte, they have been adopted as the most active and stable electrocatalyst for the efficient production of $H_2$ through HER.[13] However, their high cost and low abundance for the commercial purpose bring constraints over their potential applications to meet the energy demand and have extensively motivated them to search for better and earth-abundant HER catalysts.[14] Hence, developing a cost-effective, earth abundant, low Gibbs free energy change (ΔG) during the reaction barriers of HER electrocatalyst and lower value of the Tafel slope is paramount important for the scalable production of clean and sustainable $H_2$ energy.[9,15]

On the other side, the two-dimensional (2D) monolayer structure of transition metal dichalcogenides (TMDs) with their unique properties such as corrosion stability, tunability in the electronic properties and the possibility of defect engineering have brought an intense research in the scientific community for various applications.[16] Earth-abundant TMDs have the chemical formula of $MX_2$, where M is a transition metal atom such as Mo, W etc. and X is a chalcogen atom such as S, Se and Te. Different 2D TMDs such as molybdenum disulfide ($MoS_2$), tungsten disulfide ($WS_2$), dopped-$MoS_2$, etc., have shown HER catalytic performance with low overpotentials. Therefore, they are a promising alternative to noble Pt metal-based electrocatalysts.[16] Hinnemann et al.[17] performed a DFT[18–22] study on the 2D monolayer $MoS_2$ TMD material and found that it has HER catalytic effect because of its nearly thermoneutral hydrogen adsorption energy. Later, Jaramillo et al.[23] experimentally synthesized $MoS_2$ nano particles on Au (111) substrate by physical vapor deposition process and narrated the linear dependent of HER activity with the length of exposed edge of crystalline $MoS_2$. More specifically, they concluded that the edges of $MoS_2$ is indeed active for HER and the activity is directly proportional to the number of available active sites. Their study found that only one atom in four metal atoms at the active edges is truly responsible for $H_2$ evolution. The reaction barriers during the HER process are the most significant feature to determine the $H_2$ evolution reaction rates. Huang et al.[14] performed a DFT study on the Mo-edges ($10\bar{1}0$) of the 2D monolayer $MoS_2$ TMD material to explore the HER mechanism, possible reaction pathway and the reaction barriers by using a non-periodic cluster model system $Mo_{10}S_{21}$. Their computational study found that the Volmer-Heyrovsky mechanism is the thermodynamically favorable and convenient reaction pathway for $H_2$ formation i.e. HER process.[14] The activation energy barrier (the free energy barrier, ΔG) for this pathway during HER was found to be 17.9 kcal/mol in the solvent phase at the Poisson-Boltzmann level of theory which was in good agreement with the experimental barrier value of 19.9 kcal/mol.[14] Lowering the reaction barrier corresponding to the free energy change between the metal hydride and the positively charged proton to result $H_2$ efficiently has vital importance to develop an efficient electrocatalyst for effective HER. In this aspect, recently, Lie et al.[9] performed joint experimental and computational studies of the 2D monolayer pristine $MoS_2$, pristine tungsten disulfide ($WS_2$) and their hybrid $W_xMo_{1-x}S_2$ alloys (i.e. W-dopped $MoS_2$ alloys) synthesized over reduced graphene oxide (rGO) with the help of wet chemical process at the low temperature. They found that the $W_{0.4}Mo_{0.6}S_2$/rGO heterostructure alloy has the lowest energy barriers during both Volmer and Heyrovsky reaction steps among the pristine $WS_2$, $MoS_2$, and their other hybrid alloys $W_xMo_{1-x}S_2$. Their non-periodic DFT calculations found that the reaction energy barrier in the solvent phase during the Volmer reaction step ($H^\bullet$-migration) has the value about 11.9 kcal/mol when the HER occurs on the surfaces of the 2D monolayer $W_{0.4}Mo_{0.6}S_2$ alloy. Similarly, the Heyrovsky reaction barrier in the solvent phase of the same $W_{0.4}Mo_{0.6}S_2$ alloy is about 13.3 kcal/mol, which was much less than the pristine 2D monolayer $MoS_2$ and $WS_2$ TMD. These low energy barriers of the 2D monolayer $W_{0.4}Mo_{0.6}S_2$ alloy have become a motivation for the search of other 2D TMDs which can have further lower activation energy barriers during the hydrogen ion adsorption in the Volmer step, $H^\bullet$-migration and $H_2$ formation during either the Tafel or Heyrovsky step. Unfortunately, the 2D monolayer tungsten diselenide ($WSe_2$) has been ignored because of its lower electronic conductivity. However, due to its high activity, ultra-low thermal conductivity (0.05 W m$^{-1}$ K$^{-1}$), relatively lower band gap (~1.6 eV) and earth abundance, the 2D pristine $WSe_2$ has been identified as an efficient electrocatalyst for HER although it has been less explored for HER catalytic performance compared to other chalcogens containing TMDs.[24] The 2D monolayer pristine $WSe_2$ TMD has three atom layers (Se-W-Se) in which the central layer of transition metal tungsten (W) is sandwiched in between two layers of selenium (Se) chalcogens. In the 2D monolayer $WSe_2$, each W atom in the middle basal plane is bonded with three Se atoms at the upper plane and three Se atoms at the lower plane through the covalent bonding as shown in Figure 1. In other words, one Se atom at the top and bottom layer of the 2D monolayer $WSe_2$ is bonded with three W atoms at the middle basal plane through covalent bonding. A van der Waals (vdW) force separates different 2D $WSe_2$ layers of a bulk $WSe_2$ with Se terminating the inert basal plane (001).[16,25–28] Whereas only the edges of the 2D monolayer $WSe_2$ are catalytically active sites for HER.[26]

The HER catalytic performance of any kind of 2D TMDs like 2D $WSe_2$ depends upon the density of exposed active edge sites[26], hydrogen adsorption free energy ($\Delta G_H$), turn over frequency (TOF)[2], overpotential, Tafel slope, activation energy barriers, and so on. Lowering the activation barrier energies through hydrogen adsorption and $H_2$ formation of an electrocatalyst has a paramount importance during the HER process. In this regard, we have theoretically and computationally investigated the HER catalytic activity of the pristine 2D monolayer $WSe_2$ TMD using hybrid density functional theory (DFT)[18–20] methods to determine the reaction pathway for HER on the W-edge ($10\bar{1}0$) of the $WSe_2$. Here, we computationally designed 2D monolayer structure of the pristine $WSe_2$ TMD material, and studied the electronic band structure, total density of states (DOS) and electronic band gap with a potential application in electrochemical water splitting reactions via HER. In this work, we employed first principles-based hybrid periodic DFT methods[18–22] with van der Waals (vdW) dispersion corrections (i.e., Grimmes' -D3 dispersion corrections)[20,29–32] to calculate the equilibrium geometries, structures, band structure, electronic band gap ($E_g$) and total DOS to predict the electronic/material properties of the 2D monolayer $WSe_2$. We found that this is an excellent material for $H_2$ evolution with high electrocatalytic performance. It was found that both the exposed Se-edge ($\bar{1}010$) and W-edge ($10\bar{1}0$) edges of the $WSe_2$ are catalytic active for HER, and the (001) basal planes of the Se−W−Se tri-layer of this material are exposed surfaces. We have described the W-edges of the 2D monolayer $WSe_2$ using a finite non-periodic molecular cluster model of $W_{10}Se_{21}$, which enables to integrate of DFT accurately for calculating the reaction barriers while describing the solvation effect of water as solvent by considering the Polarization Continuum Model (PCM). This theoretical approach determined the two-electron ($2e^-$) transfer chemical mechanism for HER through Volmer-Heyrovsky and Volmer-Tafel reaction pathways by estimating the changes of Gibbs free energy i.e., relative free energy ($\Delta G$) (for the activation barriers and reaction intermediates), Turn Over Frequency (TOF) and electronic structure and properties. A comparative description of the 2D monolayer $WSe_2$ toward HER has been given with other reported 2D TMDs and their alloys. The present computational DFT study of the pristine 2D monolayer $WSe_2$ shows that it has very low activation energy barriers during the Volmer step for the adsorbed hydrogen migration ($H^\bullet$-migration) as well as the $H_2$ formation during both the Heyrovsky and Tafel reaction steps. The activation energy barriers of this $WSe_2$ material of interest during the adsorbed hydrogen migration and molecular hydrogen formation are less than other previously reported TMDs like $MoS_2$, $WS_2$, $W_xMo_{1-x}S_2$ alloys, etc. Both the Volmer-Heyrovsky and the Volmer-Tafel reaction mechanisms have been considered here and both the pathway have very low activation energy barriers in both the solvent and gas phases found in the present computational study. The lowest activation energy barriers of the pristine 2D monolayer $WSe_2$ compared with other previously stated 2D TMDs enables it to use as a potential candidate for low-cost and noble metal-free electrocatalyst for efficient production of $H_2$ through HER.

## 2. METHODOLOGY AND COMPUTATIONAL DETAILS

The computational methods along with other parameters used during the theoretical study play a vital role to determine electrocatalytic activities for efficient HER. In the present study, a periodic 2D monolayer of the $WSe_2$ was computationally designed to investigate its structural and electronic properties with the aid of first principle-based hybrid DFT method. Similarly, a non-periodic finite molecular cluster model $W_{10}Se_{21}$, corresponding to the 2D monolayer $WSe_2$ TMD was established to study the HER mechanism at the active edges through the DFT calculations. Further discussion of the periodic and non-periodic systems is explained in detailed as follow.

### 2.1. Periodic DFT Calculations

The first principle based periodic hybrid density functional theory (DFT) method was applied to obtain a symmetric 2D monolayer structure of the $WSe_2$. The equilibrium geometry with optimized lattice parameters of the symmetric 2D monolayer $WSe_2$ was obtained by using the first principle based *B3LYP-D3* (Becke, 3-parameter, Lee-Yang-Parr with Grimme's-D3 dispersion correction)[24-27,33] method implemented in *ab-initio* based CRYSTAL17 suit code.[34–39] This *B3LYP-D3* DFT method suffers less or no spin contaminations effect compared to the other post-Hartree-Fock (HF) methods which helps to provide an excellent calculations for geometry, energy and electron density.[22,33,40–44] Semi-empirical Grimme's 3rd order dispersion correction (Grimme's-D3) is taken account to encounter the non-bonding weak van del Waals (vdW) interactions among different layers and atoms to achieve a fine equilibrium geometry.[45–48] Triple ζ valence polarization (TZVP) quality Gaussian types of basis set with Gaussian types of atomic orbitals (GTO) has been used for the Se atoms and Gaussian basis sets of W_cora_1996 has been used for the W atoms in the present computations.[49–52] The convergence criteria between two consecutive iteration steps for energy, force and electron density calculations was set with a threshold value of $10^{-7}$ a.u.[53,54] In other words, a threshold value of $10^{-7}$ a.u. were used for the convergence of forces, energy, and electron density for all cases. The periodicity in the z-direction of the crystal structure was ignored by keeping the height of unit cell around ~500 Å i.e., the vacuum region of approximately 500 Å was considered in the present calculations to accommodate the vacuum environment. The vacuum region of 500 Å was set to avoid the interlayer interaction between two consecutive layers in z-direction implemented in CRYSTAL17 suite code.[34-39] VESTA visualization software is used for the provision of analyzation and graphical representation of the periodic optimized structures.[55] The equilibrium periodic 2D monolayer structure of the $WSe_2$ material is shown in the Figure 1. The optimized lattice parameters (i.e., lattice constants, symmetry, and atomic positions) of the 2D monolayer $WSe_2$ were further utilized to calculate its electronic properties such as electronic band structure and total electron density of states (DOS). The k-mesh grid is sampled on 15x15x1 Monkhorst-pack for all the integrations of the first Brillouin zone during all the computations with a resolution of around $2\pi \times 1/60$ Å$^{-1}$ for both the optimization and material properties calculations.

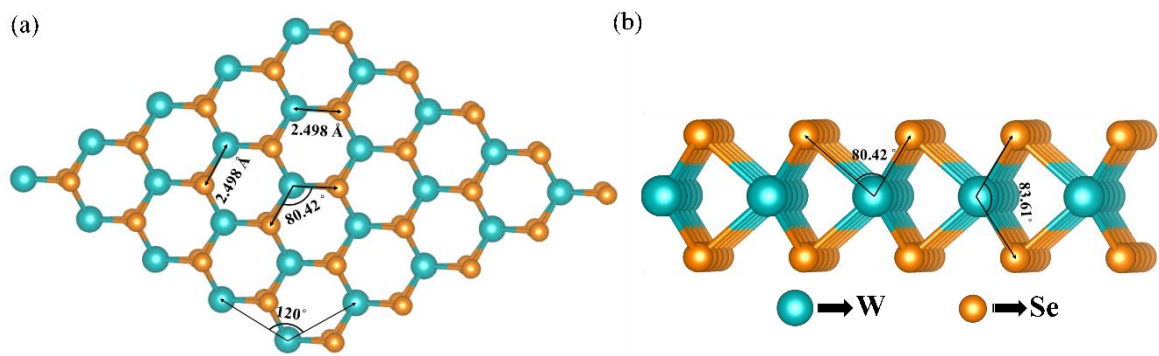

**Figure 1.** (a) Top and (b) view of the equilibrium structure of the periodic 2D monolayer $WSe_2$ material are shown here.

## 2.2. Non-periodic DFT Calculations

A non-periodic finite molecular cluster model $W_{10}Se_{21}$ of the 2D $WSe_2$ has been constructed to explore the HER mechanism on the active surface of the 2D monolayer $WSe_2$ TMD. Figure 2a shows how to extract the triangular non-periodic finite molecule cluster model from the periodic 2D monolayer $WSe_2$ to expose its either W-edges or Se-edges. The upper black dotted horizontal line represents the termination along W-edge ($10\bar{1}0$) and the bottom black dotted horizontal line denotes the termination along Se-edge ($\bar{1}010$) as shown in Figure 2a. The straight triangle represents the non-periodic finite cluster model terminated along the Se-edge ($\bar{1}010$) and the upside-down triangle corresponds to the non-periodic cluster model terminated along the W-edge ($10\bar{1}0$). Figures 2b-2d signifies the molecular cluster model containing 10 central tungsten (W) metal atoms in a single triangular plane and 21 selenium (Se) atoms (9 Se atoms in the central W plane and 6-6 Se atoms at the top and bottom plane of the central W plane).[9] The logic behind developing such finite molecular cluster is that; (i) it replicates our parental Se-W-Se tri-layer of the pristine 2D monolayer WSe$_2$ with the catalytically inert basal plane (001) and catalytically active W-edges (10$\bar{1}$0) and Se-edges ($\bar{1}$010); (ii) it provides feasibility of doping engineering as the W metal atoms can be substitutionally doped with other atoms by 10% variation of each W atom substitution and the same procedure can be adopted for the Se atoms as well, and (iii) this finite molecular cluster provides a viability of introducing protons (H$^+$) and electrons (e$^-$) separately which are essential to explore free energies as a function of electrochemical potential and pH values.[9,14] A non-periodic DFT method has been used for all the theoretical calculations to explore the HER process on the surfaces of the pristine 2D monolayer WSe$_2$ material. In the finite molecular cluster model, each W atom in the basal plane (001) has +4 oxidation state and each of them creates six bonding with six adjacent Se atoms (i.e., three Se atoms at the lower plane and three Se atoms at the upper plane). Due to this configuration, a stabilized structure is resulted in which each W-Se bonding have a 4/6=2/3 electron contributions in the inert basal plane. The stabilization of the molecular cluster model also can be understood from the oxidation state of the Se atoms in the basal plane. Each Se atom has -2 oxidation state and creates bonding with 3 W atoms that gives a contribution of 2/3 electrons towards each W-Se bonding in the basal plane. Again, the edges of the molecular cluster model are stabilized with the 2 local electron W-Se bonds with a single electron contribution towards four W-Se bonds in the basal plane as shown in the Figure 2. This 14/3 {i.e., (2×1) + [4× (2/3)]} contribution of electrons towards the W-Se bonds of the edge W atom is satisfied with the $d^2$ configuration of one W atom and $d^1$ configuration of two W atoms at the edges. With this configuration, a stabilized molecular cluster model with the periodicity 3 is achieved that derives the molecular cluster model having three edges without any unsatisfied valency. Thus, we considered a molecular cluster W$_{10}$Se$_{21}$ model system (noted by [WSe$_2$]) to represent the Se-terminated W-edges on the surfaces of 2D monolayer WSe$_2$ TMD shown in Figure 2, and this W$_{10}$Se$_{21}$ molecular cluster model system is good enough to explain the HER process.

All the required electronic and thermodynamic properties of the different steps involved in this HER mechanism have been calculated by using Minnesota 2006 local functional (*M06-L*)[56] DFT method. *M06-L* is a local spin-specific kinetic energy density ($\tau_\sigma(\mathbf{r})$, where σ = α, β and **r** = point in real space) generalized gradient approximation (GGA) method known as Minnesota DFT method.

$$\tau_\sigma(\mathbf{r}) = \frac{1}{2}\sum_{i=1}^{n_\sigma}|\nabla\psi_{i\sigma}(\boldsymbol{r})|^2$$

Where, $\psi_{i\sigma}$ represents spatial part of an occupied Kohn-Sham (KS) spin orbital, $n_\sigma$ is the number of occupied spin orbitals of the corresponding spin σ.[57] The significances of using *M06-L* local functionals are; (I) they are computational cost effective compared to non-local functionals for the molecule or the system containing large number of atoms considering density-fitting algorithms using Gaussian basis sets.[58–60] (II) This M06-L method gives more accurate and reliable energy barriers and bond energies in the case of reaction mechanism for the systems containing transition metals (TMs) like Mo, W etc., in which density-based exchange functionals describe static correction much better than the HF exchange.[61,62] DFT is widely used for large systems containing TMs as it describes electron correlation effects and have advantage over the post HF methods.[9,22,30,63–66] Double-ζ Pople-type Gaussian basis set 6-31+G$^{**}$ has been used for the selenium (Se), hydrogen (H) and oxygen (O) atoms, and LANL2DZ (Los Alamos National Laboratory 2 double-ζ) basis set with Effective Core Potentials (ECPs) for the W atoms.[67-69] The ECPs were used to replace the inner core electrons of the W atoms in the present calculations.[67–69] A harmonic vibrational frequency analysis has been carried out to confirm the stable minima of the equilibrium geometries with the successful elimination of imaginary frequencies and to locate the transition states (TSs) during HER. Gaussian16 Suite code was used to obtain the equilibrium structures and TSs of the different intermediates resulted during the HER mechanism of the 2D monolayer WSe$_2$ system.[70–72] All the optimized stable geometries of different steps involved in the HER along with the equilibrium TSs are presented with the help of ChemCraft molecular visualization software.[73]

Polarization Continuum Model (PCM) analysis has been performed at the equilibrium geometry of solvent radius 1.4 Å and H$_2$O with a dielectric constant ε = 78.54 at 298.15 K temperature[74,75] to encounter the solvation effects in the present calculations.[74] Previously, it was theoretically reported that the Gibbs free energy (G) of the solvated proton (H$^+$) was obtained to be -271.86 kcal/mol.[76] Later on, with the further theoretical study, the Gibbs free energy of the H$^+$ was found to be -270.3 kcal/mol[76] which is in the well consistent with the previously reported value.[14] Tissandier at al. theoretically estimated this value by taking the summation of gas-phase Gibbs free energy (G) value of H$^+$ at 1 atm (G (H$^+$,1 atm) = H – TS = 2.5k$_B$T – T x 26.04 = - 6.3 kcal/mol) and empirical hydrogen energy (G (H$^+$, 1 atm → 1M) = - 264.0 kcal/mol). The free energy of electron by using standard hydrogen electrode (SHE) at pH=0 is given as $G(e^-) = G\left(\frac{1}{2}H_2\right) - G(H^+)$, where the Gibbs free energy of the hydrogen molecule ($G(H_2)$) for our calculation was found computationally by performing *M06-L* local functionals DFT calculations and the free energy of proton $G(H^+)$ was taken from the previously reported values.[14]

## 2.3. Theoretical Calculations and Equations:

The electrocatalytic performances of the 2D monolayer $WSe_2$ have been characterized by the computations of the changes of Gibbs free energy ($\Delta G$) for $H_2$ adsorption on the ($\bar{1}010$) Se-edges and ($10\bar{1}0$) W-edges of the TMD. The changes of free energy ($\Delta G$), enthalpy ($\Delta H$), and electronic energy ($\Delta E$) in both the gas phase and solvent phase for all the intermediates of the HER have been calculated by the following equations.

Change of free energy: $\quad \Delta G = \sum G_{Product} - \sum G_{Reactant}$

Change of free enthalpy: $\quad \Delta H = \sum H_{Product} - \sum H_{Reactant}$

Change of free electronic energy: $\quad \Delta E = \sum E_{Product} - \sum E_{Reactant}$

Where the energies for each step participating in the HER were found computationally by the following expression.

$$E = E_{DFT} + E_{ZPE} + \int C_p \, dT - TS$$

Where, $E_{DFT}$ stands for the ground state electronic energy calculated by the DFT method, $E_{ZPE}$ represents the zero-point vibrational energy, $C_p$ is the lattice specific heat capacity at constant pressure, S represents the entropy of the system and T corresponds to the temperature at the absolute scale (here, T = 298.15 K throughout our calculations).

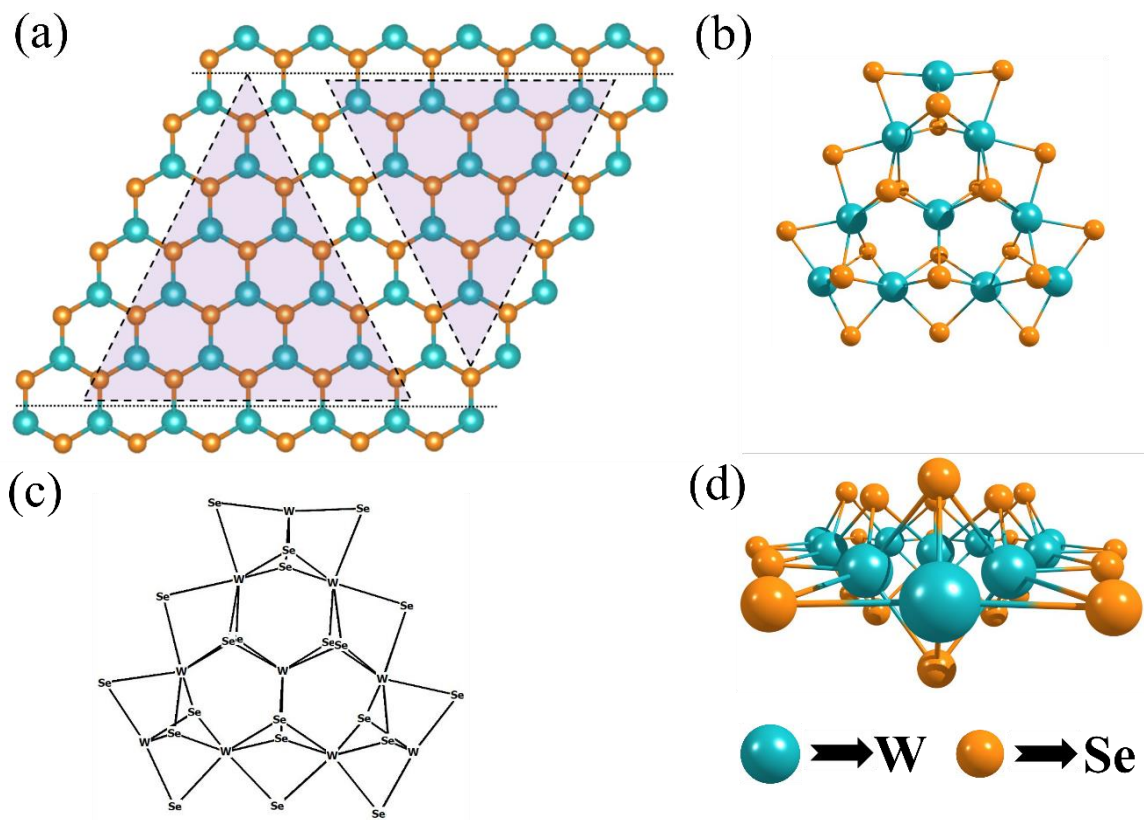

**Figure 2.** (a) Ball-and-stick style top view of the Se-W-Se tri layer periodic 2D monolayer WSe$_2$, the upper and lower black dotted horizontal lines represent the terminations along W and Se-edges, respectively. The purple colored up straight triangle represents the non-periodic molecular cluster with active Se-edge ($\bar{1}010$) termination and the upside-down purple triangle represents the non-periodic molecular cluster with active W-edge ($10\bar{1}0$) termination, (b) Top view of the designated W$_{10}$Se$_{21}$ non-periodic triangular molecular cluster of the 2D monolayer WSe$_2$, (c) Schematic representation of molecular cluster model system, (d) Side view of the chosen W$_{10}$Se$_{21}$ non-periodic triangular molecular cluster of the 2D monolayer WSe$_2$ TMD are shown here.

## 3. RESULTS AND DISCUSSION

### 3.1. Equilibrium periodic 2D monolayer structure calculations:

To acquire the fundamental understanding of the equilibrium geometry of the pristine 2D monolayer WSe$_2$ material, we have computationally designed a 2D monolayer slab of the WSe$_2$ TMD which has ***P-6m$_2$*** hexagonal 2D layer symmetry (layer group number is 78). This periodic 2D slab corresponds to the periodic symmetry along the x and y directions representing the hexagonal 2D layer system with respect to the vacuum (w.r.t. vac.) space along z-axis. In other words, there is no symmetry along z-axis. The hybrid periodic DFT-D calculation shows that the equilibrium lattice parameters are about $a = b = 3.225$ Å and the interfacial angle between them is $\gamma = 120°$ where all the atoms in the system experience minimum potential force of interactions. The equilibrium structure of the optimized pristine 2D monolayer WSe$_2$ is depicted in the Figure 1. The equilibrium average W-Se bond length between W and Se atoms in the WSe$_2$ is about 2.50 Å which is well harmonized with previous reported results,[59,60] and the in-plane bond angle < Se-W-Se is about 83.61° and the out-plane bond angle < Se-W-Se is found to be 80.42°. This in-plane bond angle < Se-W-Se is in well consistent with the previously observed value about 83.80°, calculated for the 3D bulk structure WSe$_2$.[77] The values of the equilibrium lattice parameters such as the lattice constants (*a* and *b*), interfacial angle γ, layer group symmetry and the equilibrium average bond length W-Se of the 2D monolayer WSe$_2$ are found to be well consistent with the previously reported values and they are summarized in Table 1.

**Table 1.** Equilibrium lattice parameters and the average bond length of optimized 2D WSe$_2$.

| Parameters | Optimized parameters | Previously reported values | References |
|---|---|---|---|
| *a* (in Å) | 3.23 | 3.32, 3.29 | 78 |
| *b* (in Å) | 3.23 | 3.32, 3.29 | 78 |
| α = β (in °) | 90 | 90 | 78 |
| γ (in °) | 120 | 120 | 78 |
| Symmetry | ***P-6m$_2$*** | ***P-6m$_2$*** | 79 |
| W-Se average bond length (in Å) | 2.50 | 2.55, 2.53 | 78 |

### 3.2. Electronic properties calculations:

The electronic properties calculations (i.e., electronic band structures, energy band gap ($E_g$), position of the Fermi level ($E_F$), and total density of states (DOSs)) have been performed to explore the materials properties with the contribution of electrons toward the electrocatalytic activity of the pristine 2D monolayer WSe$_2$ as shown in Figure 3. The present DFT-D (i.e.,

*B3LYP-D3*) calculation shows that the pristine 2D monolayer WSe$_2$ is a pure semiconductor with a band gap ($E_g$) energy about 2.39 eV as depicted in Figure 3. To obtain the electronic band structures of the pristine 2D monolayer WSe$_2$ material, total eight numbers of energy bands around the Fermi energy level have been computed and plotted in a specific direction of irreducible Brillouin zone by choosing *Γ-M-K-Γ* high symmetric points as depicted in Figure 3a, which is consisted with the original 2D layer group symmetry of the WSe$_2$ TMD material. It should be mentioned here that the electrostatic potential calculations have been included in the present computations for the 2D monolayer calculations of the WSe$_2$ TMD, i.e., the energies of both the band structures and total DOS are reported with respect to the vacuum (w.r.t. vac.). The Fermi energy level ($E_F$) was found at -6.59 eV which is very close to the top of valence band (VB), and a direct electronic band gap was at K point in the band structure calculations which was later confirmed by the total DOS calculations. Figure 3b represents the total density of states of the pristine 2D monolayer WSe$_2$ in which the maxima of the valence band (VB) and the minima of the conduction band (CB) was found at -6.59 eV and -4.20 eV, respectively. Here, the $E_F$ is near to the maxima of the VB and the electron density at the minima of the CB is about 2.39 eV far from the $E_F$ and the difference between these values is equal to the energy band gap ($E_g \sim 2.39$ eV). The ground state electronic configurations of the W and Se atoms are given by [Xe] $4f^{14}\ 5d^4\ 6s^2$ and [Ar] $4s^2\ 3d^{10}\ 4p^4$, respectively, so each W atom has four unpaired up spin electrons in the $5d_{xy}, 5d_{xz}, 5d_{yz}, 5d_{x^2-y^2}$ sub-shells and it has a vacant $5d_{z^2}$ sub-shell of the 5*d* orbital. Similarly, each Se atom has paired spins in $4p_x$ sub-shell and two single unpaired up spin electrons in $4p_y$ and $4p_z$ subshells of the 4*p* orbital. As both the W and Se atoms have unpaired spins in their respective higher orbitals, so the partial density of states (PDOS) calculations has been performed for both to understand the availability of states for occupation in each case. The partial density of states calculations of the W atom indicates that the most the significant contribution of the DOS goes towards the conduction band electron density of the total DOS. However, the contribution of the W atom towards the valence band electron density of the total DOS is comparatively less as shown in Figure 3c. From the partial density states calculation of the Se atom, it can be observed that the most of the contribution of density of states in the valence band of the total DOSs arises due to the Se atoms. Figure 3e elaborates the participation of 5*d* orbital of the W atom is mainly responsible for the DOSs calculations of the W atom in the 2D monolayer WSe$_2$. We show that the pristine 2D WSe$_2$ could be a better candidate for HER catalyst since the electronic properties calculations shows that the pristine 2D monolayer WSe$_2$ is a pure direct band gap semiconductor with an energy band gap ($E_g$) about 2.39 eV. The calculated band gap is well consistent with the earlier reported results. Even before this study, several pristine 2D TMDs have been used as an electrocatalyst for HER, such as $MoS_2$, $WS_2$, *etc*.[9,14,80] 2D monolayer $WSe_2$ may also be a suitable electrocatalyst for HER, so in the next section, we will study the HER mechanism by taking the non-periodic finite molecular cluster system of the 2D $WSe_2$ TMD. It has been experimentally proved that the pristine 2D monolayer tungsten selenide ($WSe_2$) TMD has emerged as a promising electrocatalyst for hydrogen evolution reaction. However, there was no theoretical and computational investigation about the catalytic activities of the 2D monolayer $WSe_2$ TMD to support or validate the experiments. In this study, we have investigated how the 2D monolayer $WSe_2$ is catalyzing the HER process, and it could be a promising candidate for effective HER electrocatalyst. A detailed description and analysis have been provided in the present investigations.

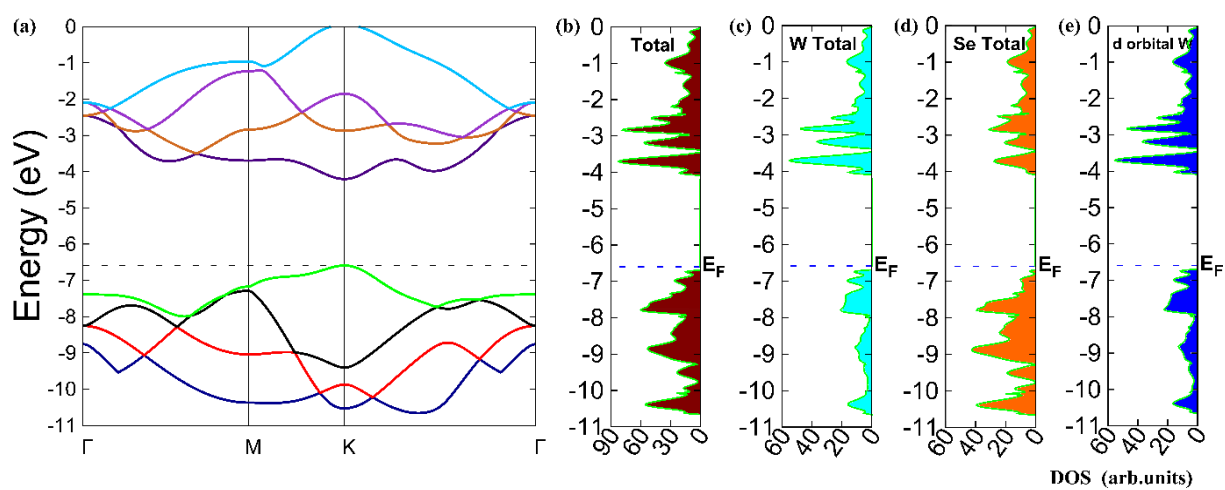

**Figure 3.** (a) Electronic band structure; (b) Total DOSs of the pristine 2D monolayer $WSe_2$; and partial density of states calculation of the (c) W atoms, (d) Se atoms, (e) *d*-orbital of the W atoms are shown here.

### 3.3. HER mechanism and thermodynamic properties calculations:

Overall electrocatalytic HER follows either Volmer-Heyrovsky or Volmer-Tafel reaction mechanisms. The study and analysis of reaction barriers during the hydrogen adsorption (Volmer reaction step), H•-migration and the hydrogen molecule desorption (either in the Heyrovsky reaction step or in the Tafel reaction step i.e., $H_2$ formation during the HER process) processes are required to explain the electrochemical activities for efficient $H_2$ evolution reactions of the electrocatalyst. In accordance with the above processes, two HER pathways are possible namely; (1) Volmer-Heyrovsky mechanism pathway in which Volmer reaction step (in short, **M + H$^+$ + e$^-$ → MH$^*_{ads}$** where protons (H$^+$) interact with the electrons, and the hydrogen atoms get adsorbed on the active region of the catalyst; where M represents the active sites of the catalyst). This reaction pathway is accompanied by the migration of H$^•$ from the Se atom site to the W transition metal atom site after the proton gets adsorbed to the active site of the catalyst. The present DFT calculations show that the first H energetically prefers to form a bond with the Se atom, as the direct hydrogen adsorption at the W-edge site is not thermodynamically favorable in the early stage of HER. Because the Se atoms are located at the edges in the 2D monolayer WSe$_2$ slab structure and the same finite non-periodic molecular cluster model system, and the basal planes of the TMD is more "inactive" than the edges. The edge sites play a major role in determining the overall activity, especially for catalysts with inactive basal planes. Generally, the edge is more reactive than the inert basal plane. However, for metal basal plane catalysts, the binding energy may be the same. The large distance between hydrogen atoms with low coverage and the rearrangement of Se atoms upon hydrogen adsorption suggests that the weakening of H bonding upon successive H adsorption is due to the limited number of basal states and the corresponding change in geometry. Therefore, a large number of edge sites has steady-state hydrogen adsorption free energies very close to -0.16 eV. The basal plane of WSe$_2$ has a limited number of active sites on the surface; however, this can be overcome by changing the surface structure. The high surface curvature of this catalyst exposes most of the edge sites, resulting in excellent catalytic activity on edges. The hydrogen binding energy ($\Delta G_H$) at the Se site in the 2D WSe$_2$ is about -0.16 eV, however, the value of $\Delta G_H$ at the W site is about +0.19 eV which is energetically less favorable. This indicates that the first H strongly prefers to bind to the Se by 0.35 eV, but the second H prefers to bind to a W instead of binding to a second Se by 0.57 eV computed by the DFT method. Recent published reports also suggest that the chalcogen site is more energetically favorable for the H adsorption than the transition metal site in other TMDs as well.[14,72,81]

In the Heyrovsky reaction step, one adsorbed hydride (H$^•$) at the transition metal site (here W) reacts with one solvated proton of the adjacent water to form one H$_2$ molecule. The second reaction mechanism follows the Volmer-Tafel reaction pathway, in which after the completion of Volmer reaction step, Tafel reaction step proceeds through one H$^•$ at the transition metal site W recombines with one H$^•$ at the Se site to form H$_2$ molecule. Our non-periodic triangular molecular cluster model $W_{10}Se_{21}$ system of the pristine 2D monolayer $WSe_2$ has the provision to introduce or remove protons ($H^+$) and electrons ($e^-$) individually to the system and to account the free energies of different intermediates resulted during the HER process. To enable the use of the most accurate DFT for reaction barriers while describing solvation effects, we used a cluster model system of the 2D monolayer $WSe_2$. This allows us to consider the introduction of protons ($H^+$) and electrons ($e^-$) separately and report free energies as a function of electrochemical potential and pH. Using molecular clusters to model a periodic system for determining reaction mechanisms allows more flexibility in the accuracy of the methods. As the complete HER process may proceed either through Volmer-Heyrovsky reaction mechanism or via Volmer-Tafel reaction mechanism, so the energy barriers of various reaction steps need to be inspected to trace the rate limiting steps. Li et al.[82] and Lie et al.[9] reported that, because of the lower hydrogen adsorption free energies of transition metal based HER electrocatalyst, Volmer-Heyrovsky reaction mechanism is the most promising reaction mechanism for $H_2$ evolution. As the 2D monolayer $WSe_2$ material has been less explored toward its electrocatalytic performance[24], so our study has been extended for both the Volmer-Heyrovsky and Volmer-Tafel proposed reaction pathways to predict the most predominant mechanism and the HER activity of the pristine 2D monolayer $WSe_2$ TMD material.

### 3.3.1 Volmer-Heyrovsky Reaction Mechanism

Volmer-Heyrovsky mechanism follows two-electron ($2\bar{e}$) transfer process, and the complete reaction processes participating in this proposed reaction pathway (when the HER occurs at the Se-terminated W-edges of the 2D monolayer $WSe_2$ material) are given schematically in the Figure 4. This multistep electrode reaction schematic representation includes the possible intermediates and the transition states (TSs) formed during the HER process. During the Volmer reaction step, the protons ($H^+$) and electrons ($e^-$) simultaneously are absorbed in the catalyst, and then the migration of the hydride ion ($H^\bullet$) occurs which is the crucial step during the $H^\bullet$-migration reaction. Similarly, in the Heyrovsky reaction step, formation of $H_2$ plays the vital role which is accomplished by the involvement of $H^\bullet$ at the transition metal site along with the requirement of one proton from the adjacent hydronium ion ($H_3O^+$). As shown in Figure 4, the complete $H_2$ evolution process accomplished with the introductions of individual electrons ($e^-$) and protons ($H^+$) in the system during HER process. It is necessary to inspect the first most stable structure of the non-periodic finite cluster model

W$_{10}$Se$_{21}$ system of the WSe$_2$ TMD along with the other structures resulted with the successive addition of each extra number of electrons and protons in order to understand the variation of free energies between the intermediates and eventually discover the possible lowest reaction barrier pathway. The detailed reaction steps involved in this proposed HER pathway are described as follow.

1. At the Standard Hydrogen Electrode (SHE) and pH = 0, the [WSe$_2$] material is the most stable state with bare neutral W-edge, which becomes the basis for our thermodynamic potential calculations. The finite molecular cluster W$_{10}$Se$_{21}$ is noted by [WSe$_2$] in short. The equilibrium structure is shown in Figure 5a.

2. To start the water splitting reactions by HER process, one electron is absorbed on the surface of the [WSe$_2$] resulting a negatively charged cluster [WSe$_2$]$^{-1}$ solvated in water with a delocalized electron on its surface. The first reduction potential of this proposed reaction pathway to obtain [WSe$_2$]$^{-1}$ from the pristine [WSe$_2$] with the introduction of a single electron is about -667.16 mV computed by the DFT method. The equilibrium geometry of the [WSe$_2$]$^{-1}$ is displayed in Figure 5b.

    Here the free energy of the electron (e$^-$) is calculated through the expression $G(e^-) = G\left(\frac{1}{2}H_2\right) - G(H^+)$. The free energy of proton, G(H$^+$) is -274.86 kcal/mol taken from previous reported value by Tissandier et al.[76], whereas the free energy of the hydrogen molecule, G(H$_2$)) is obtained by the present DFT calculations and the value is about -735.27 kcal/mol giving the free energy of electron G(e$^-$) about -92.78 kcal/mol.

3. The first hydrogen strongly prefer to bind the Se-edge rather than W atom[14], so adding a proton (H$^+$) to the Se-edge having one extra electron [WSe$_2$]$^{-1}$ results [WSe$_2$]H$_{Se}$ as an intermediate (where the subscript Se indicates that the hydrogen is bound to the Se atom) with an energy cost about 6.15 kcal/mol. The equilibrium bond length of the Se-H in the equilibrium structure of the intermediate [WSe$_2$]H$_{Se}$ is 1.47 Å, and the equilibrium geometry of the complex is displayed in Figure 5c.

4. Further adding one more electron to the [WSe$_2$]H$_{Se}$, the second reduction takes place resulting [WSe$_2$]H$_{Se}^{-1}$ (as shown in Figure 5d) with a second reduction potential about -1040 mv (i.e. -1.04 V).

5. In the next step, the hydride ion (H$^•$) adsorbed at the Se-site migrates to the adjacent W-site forming a transition state (TS) of the [WSe$_2$]H$_W^{-1}$ called as H$^•$-migration reaction step also known as Volmer transition state (TS1). This is the first transition state (TS) appeared during the HER process. A harmonic vibrational frequency analysis has been carried out to find out the TS and intrinsic reaction coordinates (IRC) calculations[42,43,83] have been performed to confirm the TS1. This TS1 is detected to have a single imaginary vibrational frequency when the H$^\bullet$ migrates from the Se-site to W-site. The equilibrium geometry of the TS1 is displayed in Figure 5e. More interestingly, the present DFT study shows that the activation energy barrier in the gas phase calculation of the H$^\bullet$ migration reaction to form the TS1 in the case of the pristine 2D monolayer WSe$_2$ is about ΔG = 2.67 kcal/mol.

The solvation effect during the HER is incorporated through the Polarization Continuum Model (PCM). In the present DFT-D calculations, we found that the energy barrier of the H$^\bullet$-migration during the TS1 formation is about 6.11 kcal/mol computed in the solvent phase i.e. water environment. This low value of the energy barrier of TS1 in both the solvent and gas phases calculation shows better hydrogen migration/adsorption than other TMDs.

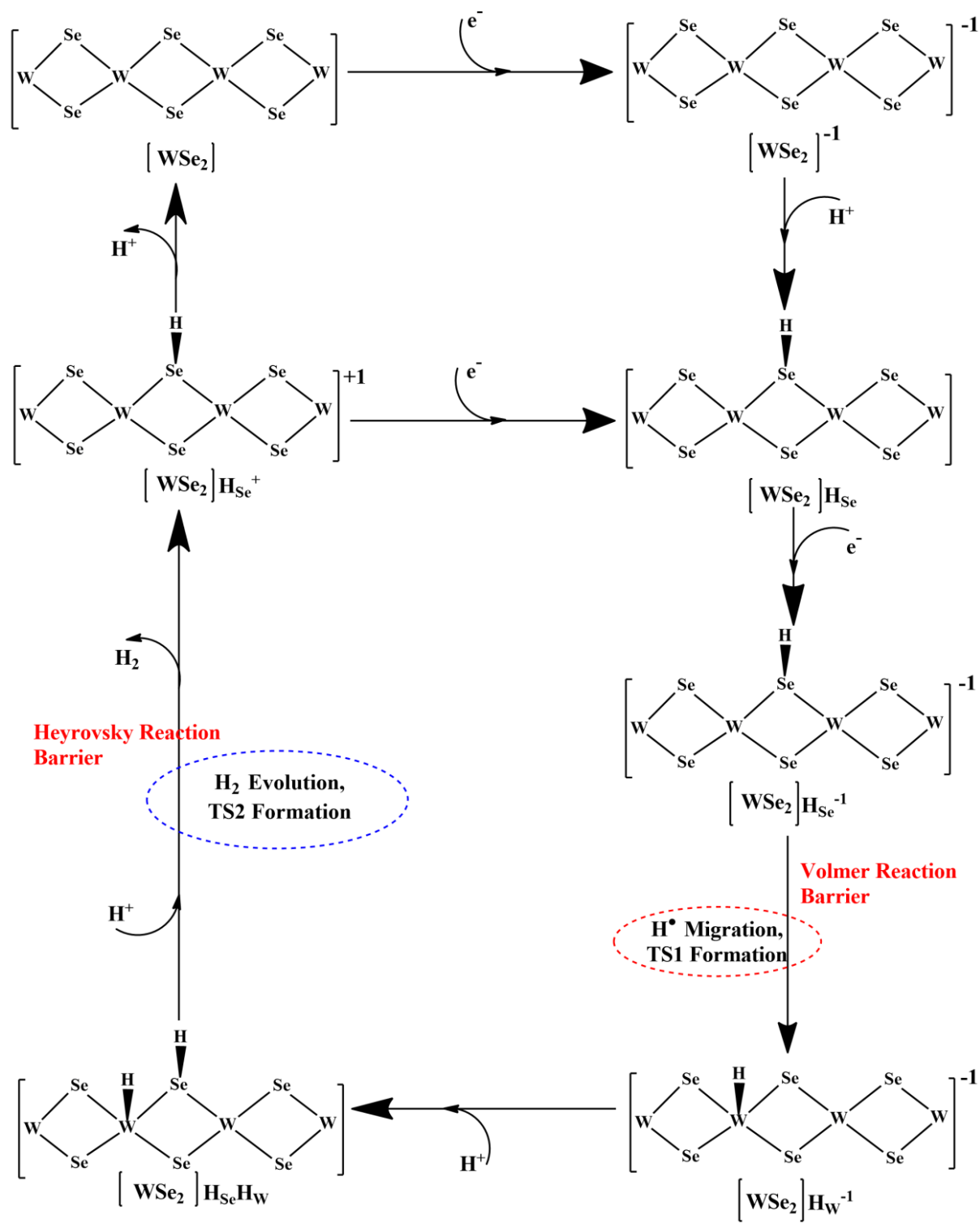

**Figure 4.** The detailed reaction mechanism of two electron transfer Volmer-Heyrovsky reaction pathway for HER on the surfaces of the 2D pristine monolayer $WSe_2$ material is displayed here.

6. The energy cost to from the [WSe$_2$]H$_W^{-1}$ complex from the TS1 is found about ΔG = -15.61 kcal/mol computed by the DFT method, and the equilibrium geometry can be found in Figure 5f. The changes of electronic energy (ΔE), relative enthalpy (ΔH) and Gibb's free energy (ΔG) during the HER process in various reaction steps computed by the M06-L DFT method are reported in Table 2.

7. One more H$^+$ from the solvent medium to the Se-site of the [WSe$_2$]H$_W^{-1}$ complex has been added to form the [WSe$_2$]H$_W$H$_{Se}$ (the subscripts W and Se at the H indicate that the one hydrogen is bound to the W atom and the other hydrogen is bound to the Se atom as depicted in Figure 5g) with an energy cost about -9.49 kcal/mol obtained by the DFT computation as shown in Table 2. The equilibrium structure of the [WSe$_2$]H$_W$H$_{Se}$ complex was found to have the equilibrium bond lengths, W-H about 1.72 Å and Se-H about 1.47 Å, respectively, computed by the same DFT method. All the optimized equilibrium structures of the intermediates and the TSs involved in the subject reaction are given in Figure 5.

8. From the [WSe$_2$]H$_W$H$_{Se}$ complex, either the Heyrovsky or the Tafel reaction process may proceed for the H$_2$ evolution. In the case of Heyrovsky reaction process, we explicitly added a hydronium water cluster (4H$_2$O_H$^+$, as shown in Figure 5h and 6) near to the active site of the W$_{10}$Se$_{21}$ (i.e. [WSe$_2$]) non periodic molecular cluster with one H$^\bullet$ at the transition metal W site and another H$^\bullet$ at the Se site resulting the [WSe$_2$]H$_W$H$_{Se}$_4H$_2$O_H$^+$ complex with an energy cost about ΔG = -11.17 kcal/mol. The equilibrium structure of the [WSe$_2$]H$_W$H$_{Se}$_4H$_2$O_H$^+$ complex is depicted in Figure 5h.

9. To proceed further HER, the second transition state has been form as known as Heyrovsky's transition state (TS2) shown in Figure 5i. The second TS known as Heyrovsky TS2 results from the [WSe$_2$]H$_W$H$_{Se}$_4H$_2$O_H$^+$ in which H$^\bullet$ from the W-site and H$^+$ from the hydronium water cluster recombine to evolve as H$_2$ and gets separated from the system as depicted in Figure 5i. The formation of H$_2$ during the reaction in TS2 is highlighted in red dotted circle in Figure 5i. The activation energy barrier of the Heyrovsky TS2 is about 6.24 kcal/mol computed in the gas phase. It should be mentioned here that in acidic media, the Volmer–Heyrovsky reaction mechanism proceeds via two proton-coupled electron transfer (PCET) steps; (1) electrosorption of a solvated proton, i.e. the Volmer reaction, followed by (2) an Eley–Rideal-type recombination of the adsorbed hydrogen (H*) and another solvated proton to form molecular hydrogen, the Heyrovsky reaction.

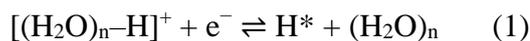

$$[(H_2O)_n–H]^+ + e^- \rightleftharpoons H^* + (H_2O)_n \quad (1)$$

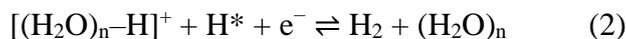

$$[(H_2O)_n–H]^+ + H^* + e^- \rightleftharpoons H_2 + (H_2O)_n \quad (2)$$

In the preceding, the solvated proton complex is denoted by $[(H_2O)_n–H]_+$, where n = 1, 2 and 4 are recognized as the hydronium ($H_3O^+$), Zundel ($H_5O_2^+$) and Eigen ($H_9O_4^+$) cations, respectively, often used to model aqueous hydronic species.

Most of the reactions for the high commercial production in the industrial level is performed in the acidic medium. So, our study has been extended to study the solvation effects during the HER by incorporating the Polarization Continuum Model (PCM) as that for TS1. The energy barrier of the TS2 during the Heyrovsky reaction step to $H_2$ formation and evolution was found about 8.41 kcal/mol in the solvent phase reported in Table 2. This TS2 is also known as Heyrovsky transition state for $H_2$ evolution.

10. After the formation of the Heyrovsky TS2, the system becomes $[WSe_2]H_{Se}^{+1}$ with the evolution of one $H_2$ molecule and four $H_2O$ molecules which is accompanied with an energy cost of -6.27 kcal.mol$^{-1}$. This is the place where $H_2$ evolves from the surface of the catalyst and again the reaction process starts from the initial steps either absorbing one electron or releasing the proton as depicted in Figure 4.

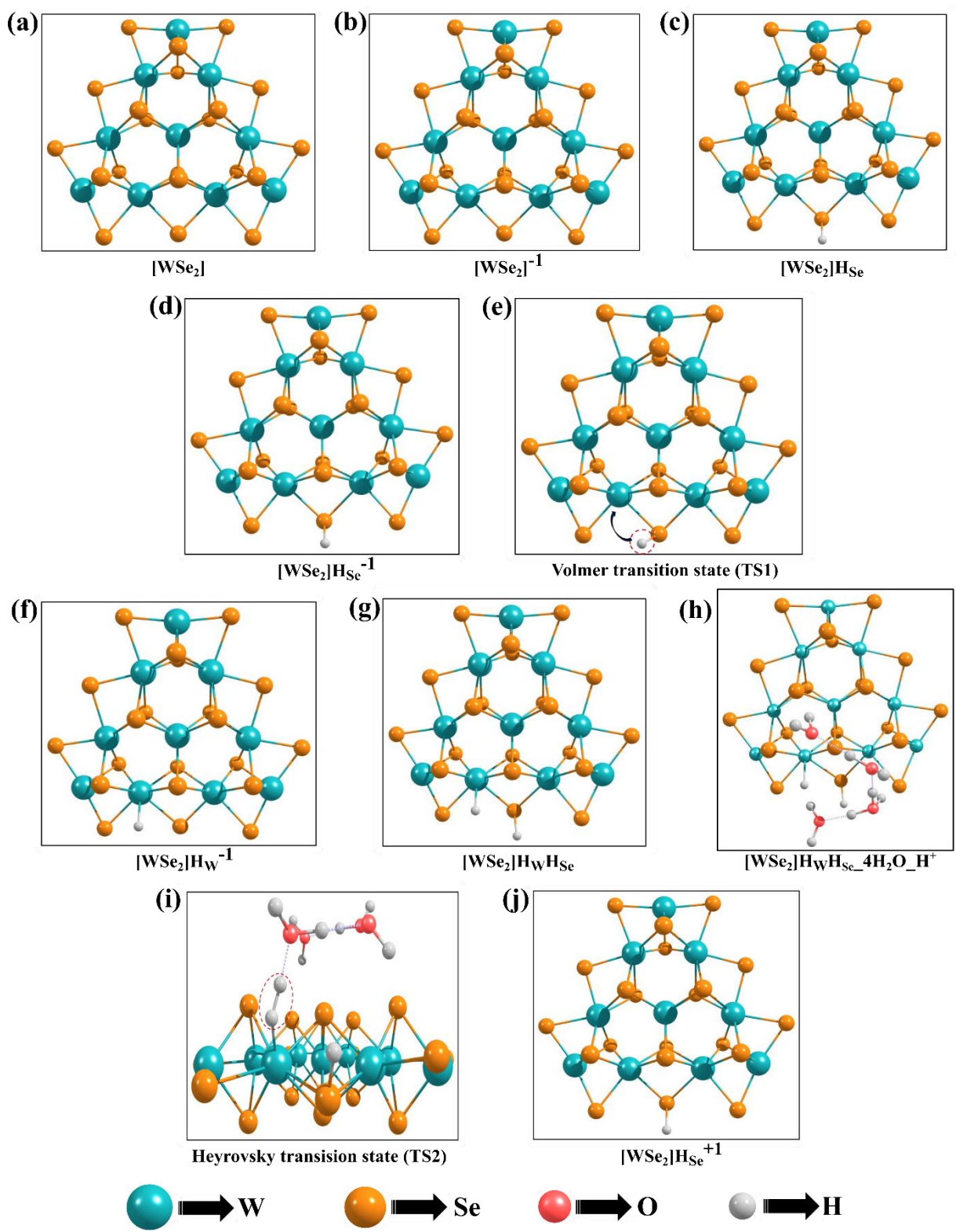

**Figure 5.** The equilibrium geometries of the non-periodic molecular cluster model of the (a) pristine $W_{10}Se_{21}$ (noted by [WSe$_2$]), (b)[WSe$_2$]$^{-1}$, (c)[WSe$_2$]H$_{Se}$, (d) [WSe$_2$]H$_{Se}$$^{-1}$ , (e) H$^\bullet$-migration or Volmer transition state (TS1) representing H$^\bullet$-migration from the Se site to the W site, (f) [WSe$_2$]H$_W$$^{-1}$, (g) [WSe$_2$]H$_W$H$_{Se}$, (h) [WSe$_2$]HWHSe_4H$_2$O_H$^+$ , (i) side-view of Heyrovsky transition state (TS2) representing the H$_2$ formation, and (j) [WSe$_2$]$^{+1}$ are shown here.

Figure 6a-6b represent the equilibrium geometry and the schematics representation of the optimized water with hydronium molecular cluster, respectively, which consist of three water molecules ($3H_2O$) and one hydronium ion ($H_3O^+$). This water hydronium cluster also can be rewritten as four water molecules and a single proton with it ($4H_2O\_H^+$). The weak H-bonds in the water molecules are highlighted by red dotted ellipse and the weak H-O bond in the hydronium cluster is highlighted by blue dotted ellipse.

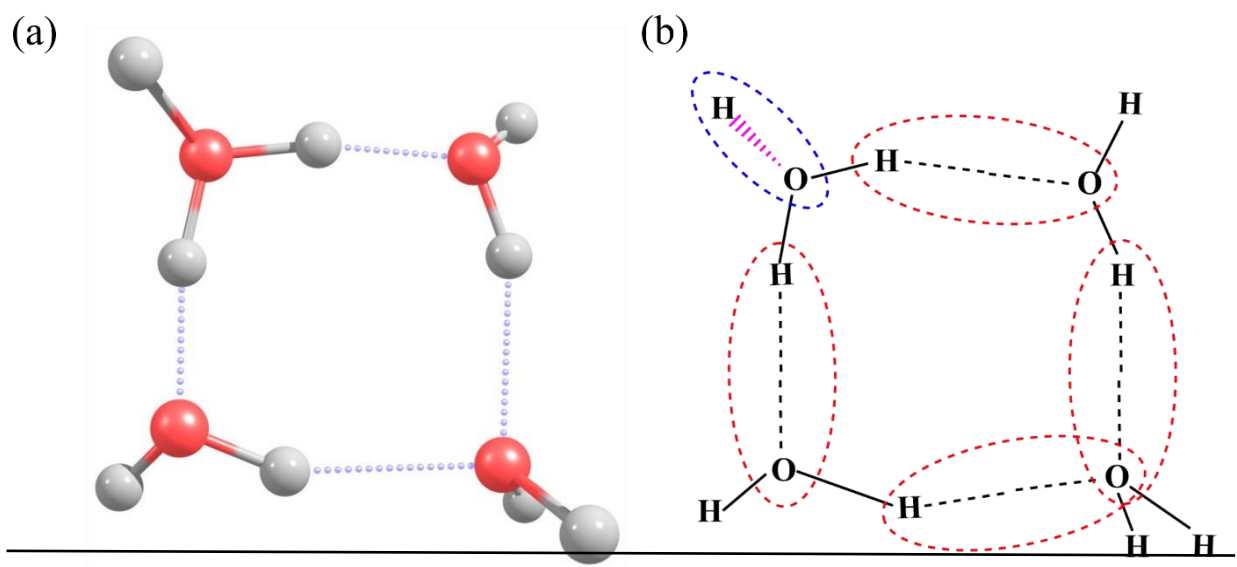

**Figure 6.** (a) Equilibrium geometry and (b) the schematic representation of the hydronium water cluster are displayed here.

As mentioned earlier that the Volmer-Heyrovsky mechanism is the most promising pathway for effective HER in the of catalysts containing transition metals[14], so our main emphasis was over the calculation of activation energy barriers for two important saddle points i.e.. Transition States (i) H•-migration or TS1 and (ii) TS2. The changes of energy (ΔE), enthalpy (ΔH) and free energy (ΔG) during different reaction intermediates as well as TSs involved in the HER followed by the Volmer-Heyrovsky reaction pathway in the gas phase calculations are summarized in Table 2.

**Table 2.** Energy changes (ΔE, ΔH and ΔG) for different intermediates and transition states (TSs) during Volmer-Heyrovsky reaction mechanism in the gas phase calculations are tabulated here. The units are expressed in kcal/mol.

| System or Reaction Intermediates/TSs | ΔE (In kcal/mol) | ΔH (In kcal/mol) | ΔG (In kcal/mol) |
|---|---|---|---|
| $WSe_2 \rightarrow [WSe_2]^{-1}$ | 15.94 | 16.13 | 15.38 |
| $[WSe_2]^{-1} \rightarrow [WSe_2]H_{Se}$ | -6.18 | -6.06 | -6.15 |
| $[WSe_2]H_{Se} \rightarrow [WSe_2]H_{Se}^{-1}$ | 23.11 | 23.16 | 24.09 |
| $[WSe_2]H_{Se}^{-1} \rightarrow H^{\bullet} - \text{migration TS1}$ | **2.79** | **2.67** | **2.67** |
| $H^{\bullet} - \text{migration TS1} \rightarrow [WSe_2]H_W^{-1}$ | -15.45 | -15.53 | -15.61 |
| $[WSe_2]H_W^{-1} \rightarrow [WSe_2]H_W H_{Se}$ | -9.48 | -9.31 | -9.49 |
| $[WSe_2]H_W H_{Se} \rightarrow [WSe_2]H_W H_{Se\_}4H_2O\_H^+$ | -24.96 | -24.76 | -11.17 |
| $[WSe_2]H_W H_{Se\_}4H_2O\_H^+ \rightarrow \text{Heyrovsky TS2}$ | **5.73** | **5.64** | **6.24** |
| $\text{Heyrovsky TS2} \rightarrow [WSe_2]H_{Se}^{+1}$ | 8.48 | 9.64 | -6.27 |

The present DFT computations found a reaction barrier about ΔG = 2.67 kcal/mol during the H$^{\bullet}$-migration reaction or TS1. Similarly, a reaction barrier about 6.24 kcal/mol has been found during the formation of TS2 in the Heyrovsky reaction process at the W-edges of the pristine 2D monolayer WSe$_2$ material (in gas phase calculation) during HER. This higher value of the Heyrovsky TS2 than the H$^{\bullet}$-migration TS1 signifies that the Heyrovsky reaction step is the rate-determining step in the Volmer-Heyrovsky reaction mechanism of the HER process in the case of the pristine 2D monolayer WSe$_2$ TMD. A potential energy surface (PES) of this Volmer-Heyrovsky reaction mechanism has been drawn and shown in Figure 7. The variations of Gibbs free energies (ΔG) changes with respect to the proceeding of the reaction steps involved in the Volmer-Heyrovsky reaction are depicted in Figure 7. As in the commercial field, most of the reactions are carried out in the form of the solution, so a solvent phase calculation of the activation barriers is also necessary to be determined. So, the solvent phase calculation has also been performed by considering the solvent effect of universal solvent

"water", and the respective reaction barriers have been noted down as explained in the above steps of the Volmer-Heyrovsky mechanism.

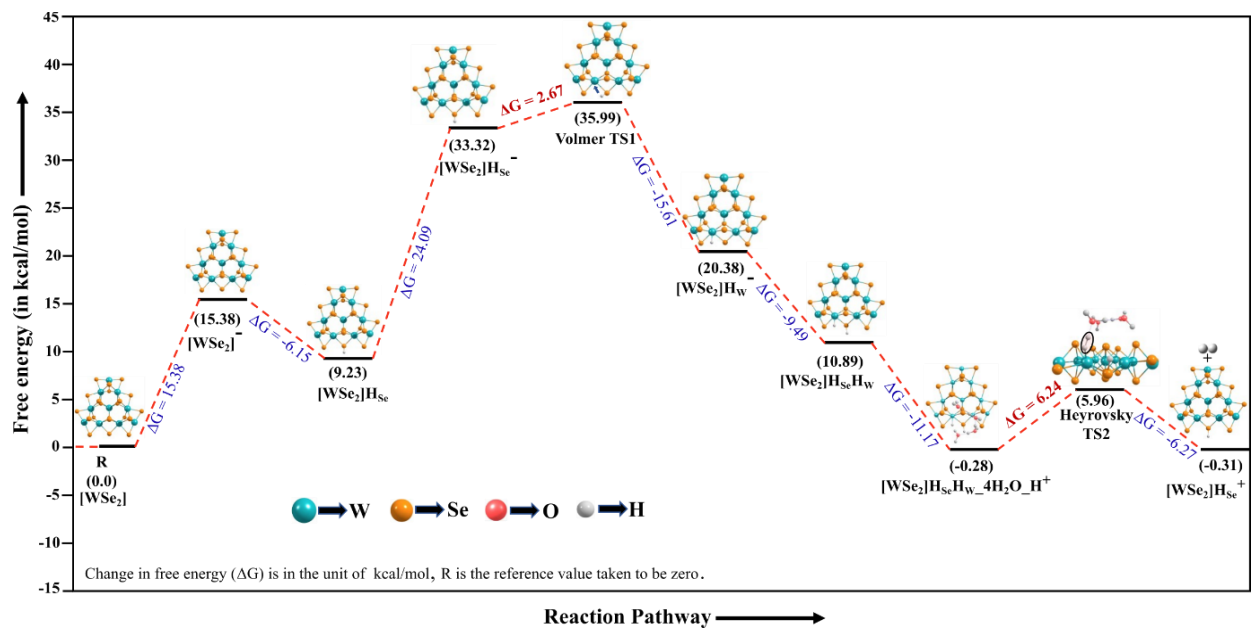

**Figure 7.** PES of the Volmer-Heyrovsky reaction mechanism during HER process at the surface of the $WSe_2$ material is depicted here. The changes of free energies ($\Delta G$) diagram are expressed in kcal/mol.

It was computationally studied that the activation energy barrier is about 17.9 kcal/mol in the solvent phase at the Poisson-Boltzmann level for the $H_2$ evolution during the Volmer-Heyrovsky reaction mechanism when the reaction occurs at the Mo-edge ($10\bar{1}0$) of the 2D monolayer pristine $MoS_2$ material.[14] The experimental value for this energy barrier has been reported to be 19.9 kcal/mol at the surfaces of the 2D $MoS_2$ material. Recently Lie et al.[9] performed a combined experimental and theoretical study on the HER efficiency of pristine $MoS_2$, $WS_2$, 2D TMDs and the heterostructures of $MoS_2$/rGO, $WS_2$/rGO and $W_xMo_{1-x}S_2$/rGO TMDs alloys. Their DFT calculations found that the reaction energy barrier of the $W_{0.4}Mo_{0.6}S_2$ developed over reduced graphene oxide heterostructure alloy has shown the lowest activation barriers compared to the pristine 2D monolayer $MoS_2$ and $WS_2$ TMDs. Their studies found an energy barrier about 11.9 kcal/mol during the H• -migration from the S site to the W transition metal site when the HER process occurs on the surfaces of the $W_{0.4}Mo_{0.6}S_2$ alloy in the solvent phase considering finite molecular cluster model system. Their theoretical calculations revealed that the $W_{0.4}Mo_{0.6}S_2$ alloy follows the Volmer-Heyrovsky reaction mechanism with the solvent phase energy barrier of about 13.3 kcal/mol during the $H_2$ evolution i.e., the Heyrovsky transition state TS2. More significantly, the present DFT calculation indicates that the 2D pristine monolayer $WSe_2$ material has the lower reaction energy barriers for both the $H^\bullet$-migration step and $H_2$ formation during the Heyrovsky reaction step in the Volmer-Heyrovsky mechanism pathway compared to other previously reported 2D TMDs materials like pristine 2D monolayer $WS_2$, $MoS_2$ and hybrid $W_xMo_{1-x}S_2$ alloys.[9] Interestingly, the present DFT study shows that the activation energy barriers during the $H^\bullet$-migration of Volmer reaction on the surfaces of the 2D pristine $WSe_2$ material are about 2.67-6.11 kcal/mol in the gas and solvent phase calculations. Similarly, the activation energy barriers of the $H_2$ formation during the Heyrovsky reaction step are about 6.24 -8.41 kcal/mol in the gas and solvent phases, respectively. These lowest activation energy barriers during $H^\bullet$ migration and $H_2$ formation (in both the gas and solvent phases computations) confirm the better HER catalytic activity of the pristine 2D monolayer $WSe_2$ than other 2D TMDs. It should be mentioned here that its electrocatalytic activity is close enough to the Pt or other noble materials-based electrocatalysts. Figure 8 represents the comparison of energy barriers of the $H^\bullet$-migration reaction during the Volmer reaction step and $H_2$ evolution in the Heyrovsky reaction step of various reported 2D TMDs along with the 2D monolayer $WSe_2$. Figure 8a-8b corresponds to the activation energy barriers of the $H^\bullet$-migration (when the $H^\bullet$ migrates from the chalcogen site to the respective transition metal site) obtained in the gas and solvent phase calculations. Similarly, Figure 8c-8d demonstrate a comparison of activation energy barriers of the $H_2$ evolution during the Heyrovsky reaction step of various previously reported 2D TMD catalysts and the pristine 2D monolayer $WSe_2$ material computed in both the gas and solvent phases. Figure 8 shows that the pristine 2D monolayer $WSe_2$ material has the lowest activation energy barriers among the previously reported 2D TMDs such as pristine $MoS_2$, $WS_2$ and hybrid $W_xMo_{1-x}S_2$ alloys. 2D monolayer $WSe_2$ (represented by dark green color bar in the 12$^{th}$ position of each bar diagram of the Figure 8) has the lowest height indicating the lowest activation energy barriers during both the gas and solvent phase calculations of both the transition states TS1 and TS2 among other 2D TMDs.

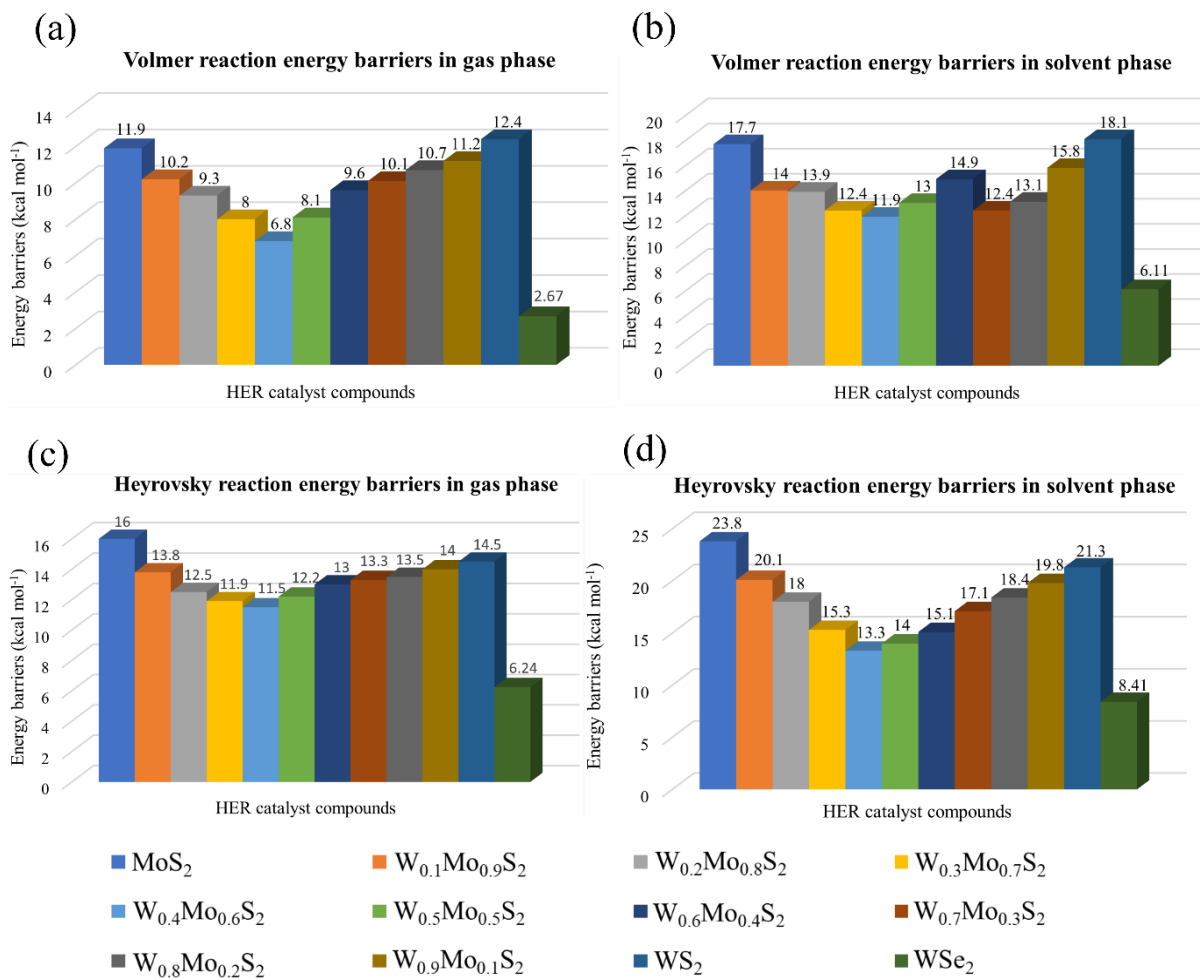

**Figure 8.** Bar-diagram and pictorial representation of activation energy barriers of the (a) gas-phase calculations of the H•-migration or Volmer TS1, (b) solvent phase calculations of H•-migration or Volmer TS1, (c) gas-phase calculations of Heyrovsky TS2, and (d) solvent phase calculations of the Heyrovsky TS2 appeared during HER process on the surfaces of various previously reported TMDs along with our system of interest are given here

### 3.3.2 Volmer-Tafel Reaction Mechanism

The Volmer-Tafel reaction mechanism is also a two-electron transfer process similar to the Volmer-Heyrovsky reaction mechanism although it is not as complex as the Volmer-Heyrovsky reaction mechanism. In the case of the Volmer-Tafel reaction mechanism, two adsorbed adjacent hydrogens on the surface of the catalyst recombine with each other to form $H_2$ as $H^{\bullet} + H^{\bullet} \rightarrow H_2$ and it does not require any further solvated proton like Volmer-Heyrovsky mechanism. The complete reaction processes participating in this proposed reaction pathway are given in Figure 9.

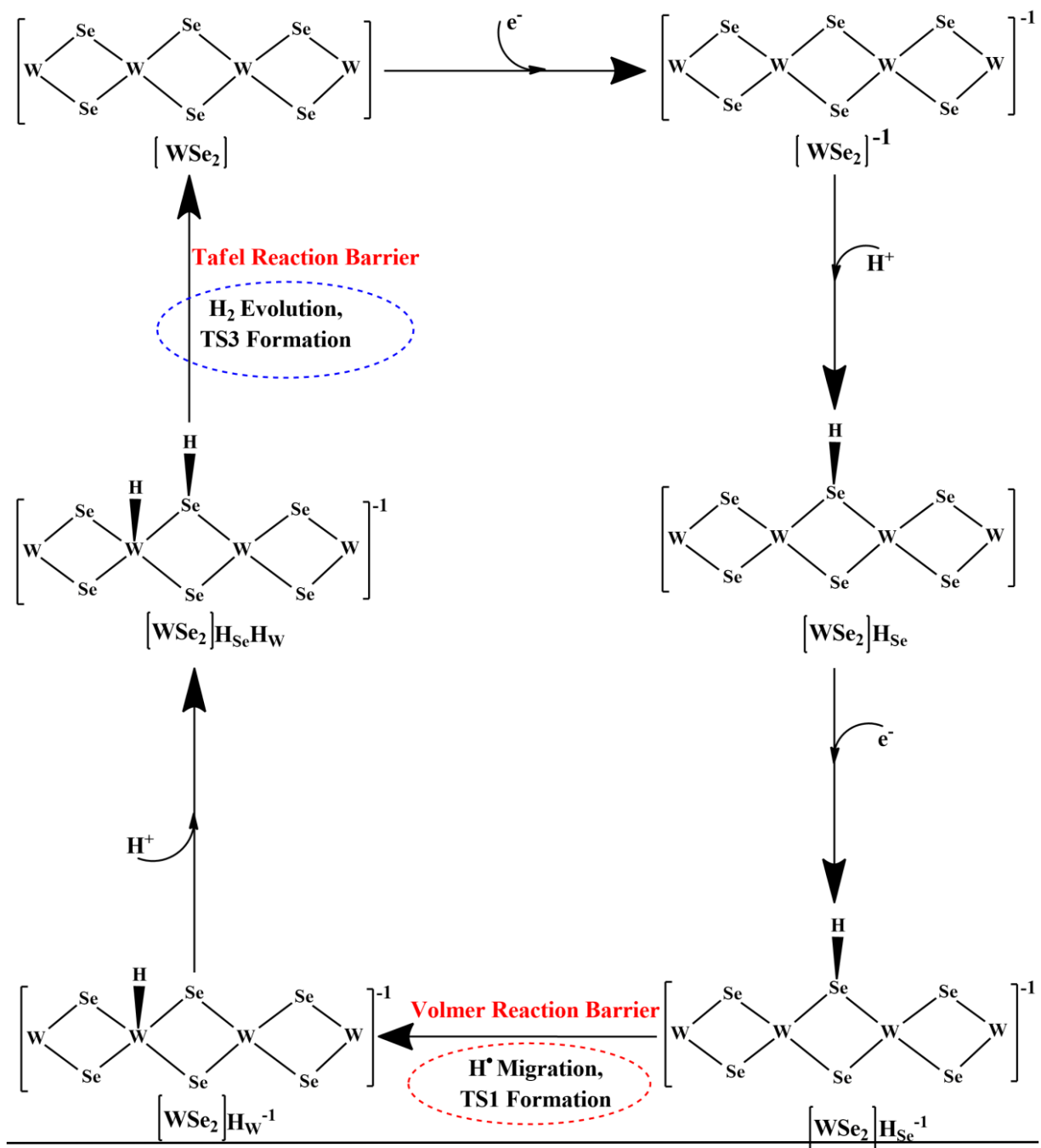

**Figure 9.** Schematic presentation of the proposed two-electron transfer, Volmer-Tafel reaction pathway for HER on the surfaces of the pristine 2D monolayer $WSe_2$ is shown here.

The detailed reaction steps involved in this proposed Volmer-Tafel reaction pathway are described below;

1. This Volmer-Tafel mechanism pathway follows the same route up to the point where the $[WSe_2]H_WH_{Se}$ intermediate results from the $[WSe_2]H_W^-$ intermediate with an energy cost of -9.49 kcal/mol with the successive addition of one proton ($H^+$) as discussed in the case of Volmer-Heyrovsky mechanism pathway.

2. In the next step, two adsorbed hydrogens (one at the W-site and the other at the Se-site) next to each other combine to form H₂, and this process is known as the Tafel reaction. The reaction phenomena of the H• + H• → H₂ gives to another TS known as the Tafel transition state (TS3), which was computationally found during the HER process. This Tafel TS3 has a single imaginary frequency with an equilibrium W-H bond length 2.21 Å, Se-H bond length 2.45 Å and the final H-H bond length about 0.78 Å. The activation energy barrier of TS3 was found to be 4.56 kcal/mol in the gas phase obtained by the same M06-L DFT method. The equilibrium geometry of the TS3 is depicted in Figure 10 along with its schematic representation. The energy barrier of the same TS3 during the Tafel reaction step for H₂ evolution was found about 6.61 kcal/mol in the solvent phase calculations obtained by the same level of theory.

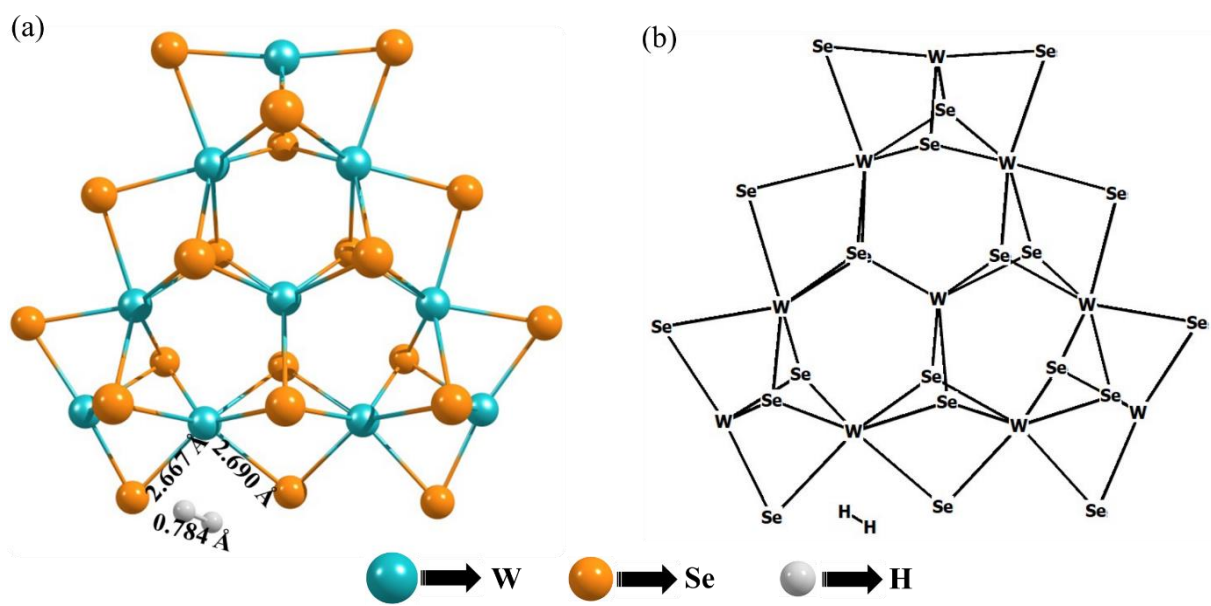

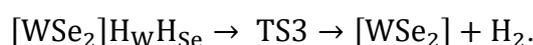

**Figure 10.** (a) Equilibrium geometry and (b) schematic representation of the Tafel transition state TS3 obtained by the M06-L DFT method are shown here.

3. After forming the TS3, the system comes to its initial step [WSe₂] along with the evolution of a single H₂, which is achieved with an energy cost of -15.45 kcal/mol and the reaction scheme can be noted by:

$$[WSe_2]H_W H_{Se} \rightarrow TS3 \rightarrow [WSe_2] + H_2.$$

The changes in energies during the various steps of the Volmer-Tafel reaction mechanism in the gas phase calculations are reported in Table 3.

**Table 3.** Energy changes (ΔE, ΔH and ΔG) for different intermediates and transition states (TSs) during Volmer-Tafel reaction mechanism are tabulated here. The units are expressed in kcal/mol.

| Reaction Intermediates | ΔE (In kcal mol$^{-1}$) | ΔH (In kcal/mol) | ΔG (In kcal/mol) |
|---|---|---|---|
| $WSe_2 \rightarrow [WSe_2]^{-1}$ | 15.94 | 16.13 | 15.38 |
| $[WSe_2]^{-1} \rightarrow [WSe_2]H_{Se}$ | -6.18 | -6.06 | -6.15 |
| $[WSe_2]H_{Se} \rightarrow [WSe_2]H_{Se}^{-1}$ | 23.11 | 23.16 | 24.09 |
| $[WSe_2]H_{Se}^{-1} \rightarrow H^{\bullet}$ − migration TS1 | **2.79** | **2.67** | **2.67** |
| $H^{\bullet}$ − migration TS1 $\rightarrow [WSe_2]H_W^{-1}$ | -15.45 | -15.53 | -15.61 |
| $[WSe_2]H_W^{-1} \rightarrow [WSe_2]H_WH_{Se}$ | -9.48 | -9.31 | -9.49 |
| $[WSe_2]H_WH_{Se} \rightarrow TS3$ | **4.59** | **4.79** | **4.56** |
| $TS3 \rightarrow WSe_2$ | -8.11 | -6.57 | -15.45 |

In the case of the pristine 2D monolayer WSe$_2$ TMD, the calculated energy barrier of the H$_2$ evolution during the Tafel reaction step of the Volmer-Tafel mechanism is much less than the other reported 2D TMDs catalysts. The energy barrier in the Tafel reaction step (i.e., TS3) is also less than the energy barrier of the Heyrovsky reaction step (i.e., TS2) for the H$_2$ evolution in both the gas and solvent phases calculations. From the gas phase calculation of the Volmer-Tafel reaction pathway of HER, we found a reaction barrier about ΔG = 2.67 kcal/mol during the H$^{\bullet}$ -migration in the TS1 and a very less reaction barrier about 4.56 kcal/mol during the TS3 Tafel reaction step. The values of these barriers are in the range of the DFT accuracy ~4 kcal/mol. This higher value of Tafel TS3 compared to the TS1 signifies that the TS3 Tafel reaction step is the rate-determining step in the Volmer-Tafel reaction mechanism of HER when the reaction takes place on the surfaces of the 2D pristine monolayer WSe$_2$. The relative variations of the Gibbs free energies with respect to the proceeding of the reaction steps involved in the gas phase calculation of the Volmer-Tafel mechanism i.e., the potential energy surfaces (PESs) are given in Figure 11. Here, the relative changes of Gibbs' free energy for all the intermediates and TSs are considered with respect to the Gibbs free energy value of the pristine [WSe$_2$] as the reference value by adopting its energy value as G([WSe$_2$]) - G([WSe$_2$])

= 0 kcal/mol. All the values of relative free energies are reported with respect to this reference value depicted in Figure 11.

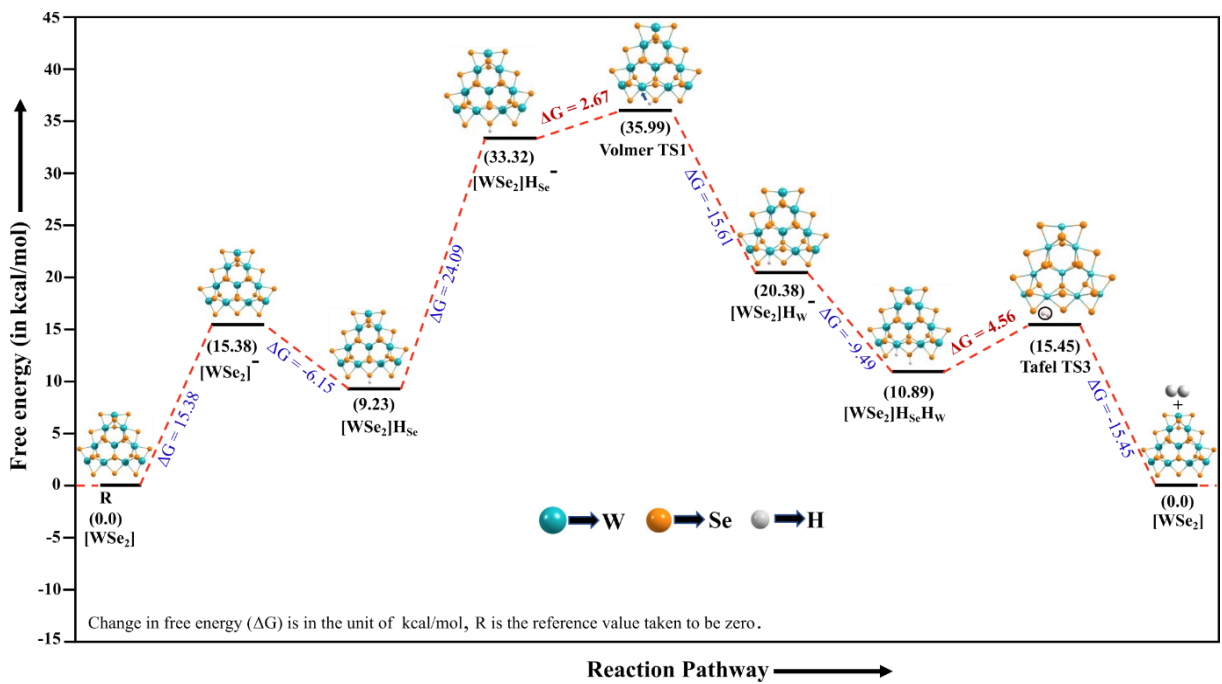

**Figure 11.** PESs of the Volmer-Tafel reaction mechanism during the HER process at the surface of the WSe$_2$ material is depicted here. The relative changes of the free energies (ΔG) diagram are expressed in kcal/mol.

In summary, the present DFT study (of the HER on the surfaces of the W$_{10}$Se$_{21}$ non-periodic finite molecular cluster model system of the pristine 2D monolayer WSe$_2$) implies that the activation reaction barrier of the TS3 in the Tafel reaction step is about 4.56 - 6.61 kcal/mol in the gas and solvent phases calculations, respectively. Here, it is noteworthy that the value of the energy barrier (ΔG) of the TS3 appeared in the Tafel reaction step (in the Volmer-Tafel reaction mechanism) is about 1.68 - 1.80 kcal/mol less than the barrier of the TS2 i.e., the Heyrovsky transition state in the Volmer-Heyrovsky reaction mechanism in both the gas and solvent phases. These values are in the range of the DFT accuracy, and these lower values of the activation energy barriers of the TS1, TS2 and TS3 in the proposed reaction mechanisms indicate that both the pathways have comparable lower energy barriers and both of them can be the assurance pathway for superior HER catalytic performance of the 2D monolayer WSe$_2$ material. These lower reaction barriers signify that the H$_2$ evolution may proceed through any of these pathways with a comparative reaction barrier like the noble metals-based catalysts. The numerical values of the reaction barriers corresponding to different TSs in both reaction pathways are depicted in Table 4 and are graphically presented in Figure 12 to provide a better comparative visualization of the reaction barriers when the HER takes place on the surfaces of the 2D $WSe_2$ TMD.

**Table 4.** Reaction barriers of different TSs in the HER mechanism of the pristine 2D monolayer $WSe_2$ material computed both in the solvent and gas phases are given here.

| Activation Barriers | ΔG (kcal/mol) In gas phase | ΔG (kcal/mol) In solvent phase |
|---|---|---|
| TS1 for H•-migration | 2.67 | 6.11 |
| TS2 for the Heyrovsky reaction barrier | 6.24 | 8.41 |
| TS3 for the Tafel reaction barrier | 4.56 | 6.61 |

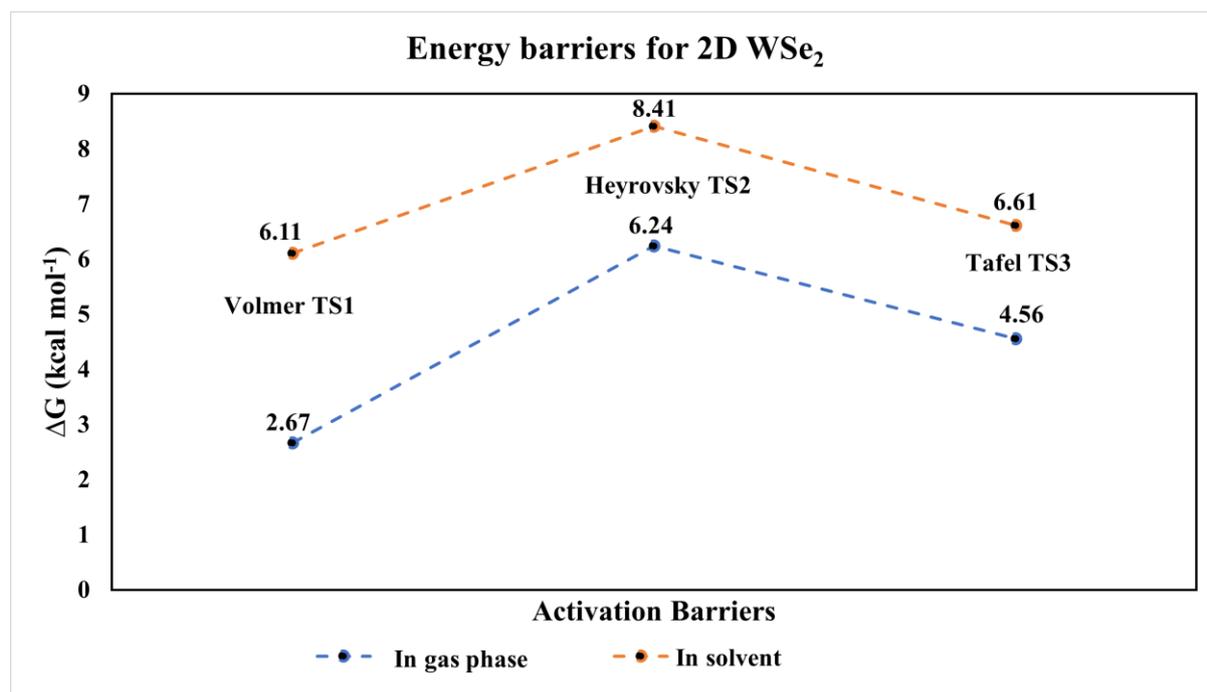

**Figure 12.** Graphical representation of the reaction barriers for HER on the surfaces of the pristine 2D monolayer $WSe_2$ material is depicted here.

## 3.4 Highest Occupied Molecular Orbital (HOMO) and Lowest Unoccupied Molecular Orbital (LUMO) Calculations:

The highest occupied molecular orbital (HOMO) and lowest unoccupied molecular orbital (LUMO) calculations have been performed at the equilibrium structures of the transition states to explain the HER process. The role of the atomic orbitals during the formation of TS1 (in the H·-migration reaction step), TS2 and TS3 ($H_2$ evolution process during either the Heyrovsky or Tafel reaction steps) can be realized with the visualizations of the HOMO and LUMO calculations of their respective TS structures. The HOMO and LUMO structures of the transition states TS1, TS2 and TS3 are shown in Figure 13. These HOMO and LUMO calculations represent the molecular/atomic orbitals overlapping of multi-electron wavefunction densities. HOMO and LUMO structures of the TS1 during the H·-migration are shown in Figure 13a-13d, in which the shifting of the electronic wavefunction density of the H· takes place from the Se site to the W metal atom site. The red color bubble represents the in-phase overlapping of the electron cloud of the 1s orbital of H· atom towards the electron cloud of the 5d orbital of the W atom during the H·-migration reaction at the TS1. The H·-migration in the HOMO and LUMO calculation of the TS1 is highlighted by a black dotted circle as shown in the Figure 13a and13d. The energy values of the HOMO and LUMO of the TS1 are $E_{HOMO}$= -7.14x$10^{-2}$ eV and $E_{LUMO}$= -6.94x$10^{-2}$ eV, respectively, with a HOMO-LUMO gap about $E_{GAP}$=$E_{LUMO}$-$E_{HOMO}$= 2.00x$10^{-3}$ eV corresponding to a photon frequency 4.91x$10^{11}$ Hz. The HOMO-LUMO calculation indicates that a wavelength $\lambda$=611.25 µm is required for the transition of an electron from the HOMO to LUMO. This HOMO-LUMO gap is an important parameter for predicting the stability and the color of the complex in the solution.

The HOMO and LUMO structures of the TS2 during the Heyrovsky reaction step of the $H_2$ evolution are depicted in Figure 13b and 13e. The formation of $H_2$ in a steady-state takes place due to stabilization of the system by better atomic orbitals overlapping of the 5d-orbital electron clouds of the W atoms and the 1s-orbital electron clouds of the $H_2$ molecules as shown in Figure 13 b and 13e. It has been also observed that the electron wavefunction of the 1s-orbital of H· at the W atom overlaps with the $H^+$ from the adjacent hydronium ion with water cluster (4$H_2$O_$H^+$) to result an $H_2$ molecule and is high lightened by a black dotted circle as shown in the HOMO structure of the TS2 in Figure 13b. The energy values of the HOMO and LUMO of the TS2 are about -0.29 eV and -0.25 eV, respectively, with a HOMO-LUMO gap about -0.04eV. This HOMO-LUMO gap corresponds to a photon frequency of 9.19x$10^{12}$ Hz and wavelength of $\lambda$=32.65 µm which is the minimum energy required to transfer an electron from the HOMO to LUMO of TS2. Similarly, Figures 13c-13f correspond to the HOMO and LUMO structures of the Tafel TS3 during the $H_2$ formation with the recombination and formation of covalent bonding between one H• at the W site and one H• at the Se site during Tafel reaction. The energy level of the HOMO is at -0.20 eV and the energy level of the LUMO is at -0.17eV resulting in a HOMO-LUMO gap about -0.03eV in the case of Tafel TS3. The gentle orbital overlapping of molecular orbitals during H•-migration in the Volmer reaction step and $H_2$ formation in the Heyrovsky and Tafel reaction steps also reveal the excellent catalytic activity of the pristine 2D monolayer $WSe_2$ for effective HER.

In other words, a better atomic orbital overlap of the *s*-orbitals of the hydrogen atom attached with the W in the $WSe_2$ material and the water cluster ($3H_2O + H_3O^+$) seemed in the HOMO-LUMO Heyrovsky's transition state TS2 during $H_2$ formation has been observed in Figure 13. Similarly, a better atomic orbital overlap of the *s*-orbitals of the hydrogen atom with the W atom during the H•-migration at the TS1 shown in Figure 13. These better electron cloud overlap of the atomic orbitals during the H•-migration at the TS1 and the $H_2$ formation in the Heyrovsky's/Tafel's TS2/TS3 have diminished the reaction barrier during the HER process taken place at the surfaces of the 2D $WSe_2$ TMD. Therefore, it can say that this better stabilization (compared to the pristine 2D monolayer $MoS_2$ and $WS_2$ with their hybrid $W_xMo_{1-x}S_2$ alloys)[9] of the atomic orbitals in the reaction rate-limiting step TS2 for $H_2$-formation is a key for reducing the Heyrovsky's reaction barrier, thus the overall catalysis indicating a better electrocatalytic performance for $H_2$ evolution.

**Turnover frequency (TOF) and Tafel slope Calculations:**

To compare the catalytic activities of electrocatalysts, turnover frequency (TOF) corresponding to the number of $H_2$ evolved per active site per unit time is another crucial parameter. The higher value of TOF means the better active catalyst, so a catalyst with a higher value of TOF is the need of the concern. Using the transition state theory (TST), the TOF at a specific temperature is given theoretically by;

$$\text{TOF} = \left(\frac{K_B T}{h}\right) e^{\left(\frac{-\Delta G}{RT}\right)}$$

where $K_B$ = Boltzmann constant ($3.298 \times 10^{-27}$ kcal/mol), T = absolute temperature (here it is specified with 298.15 K for our cluster model calculation), h = Planck's constant ($1.584 \times 10^{-37}$ kcal sec), $\Delta G$ = energy barrier and R = universal gas constant ($1.987 \times 10^{-3}$ kcal $K^{-1}$ $mol^{-1}$). Lie et al.[9] computationally found the TOF about $2.1 \times 10^{-5}$ sec$^{-1}$ for the pristine 2D MoS$_2$, $1.5 \times 10^{-3}$ sec$^{-1}$ for the pristine 2D monolayer WS$_2$ in the solvent phase calculations. The highest value of the TOF was about $1.1 \times 10^3$ sec$^{-1}$ of the W$_{0.4}$Mo$_{0.6}$S$_2$ TMD alloy material among different W$_x$Mo$_{1-x}$S$_2$ alloys.[9] Jaramillo et al. computationally estimated the TOF value about $1.64 \times 10^{-2}$ sec$^{-1}$ corresponding to the molybdenum edge of the pristine 2D single layer MoS$_2$.[23] The present DFT study reveals that the value of TOF is about $8.86 \times 10^7$ sec$^{-1}$ when the HER takes place at the W edges of the pristine 2D monolayer WSe$_2$ TMD during the solvent phase reaction in the case of Volmer-Tafel reaction mechanism with an activation energy barrier of 6.61 kcal/mol. Similarly, the value of TOF was found to be $4.24 \times 10^6$ sec$^{-1}$ corresponding to the activation energy barrier of 8.41 kcal/mol (in the solvent phase) during the Heyrovsky reaction at the W-active edges of the pristine 2D monolayer WSe$_2$ TMD. This ultra-high value of the TOF (in the case of pristine 2D monolayer WSe$_2$) is much higher (around $3.65 \times 10^3$ times) than the value of TOF of the W$_{0.4}$Mo$_{0.6}$S$_2$ TMD alloy material during the Heyrovsky step in solvent phase HER and the same is around $8.05 \times 10^4$ times higher if we consider the solvent phase calculation of the Tafel reaction step. This high value of TOF is a significant indicator for the excellent performance of the pristine 2D monolayer WSe$_2$ for the efficient HER.

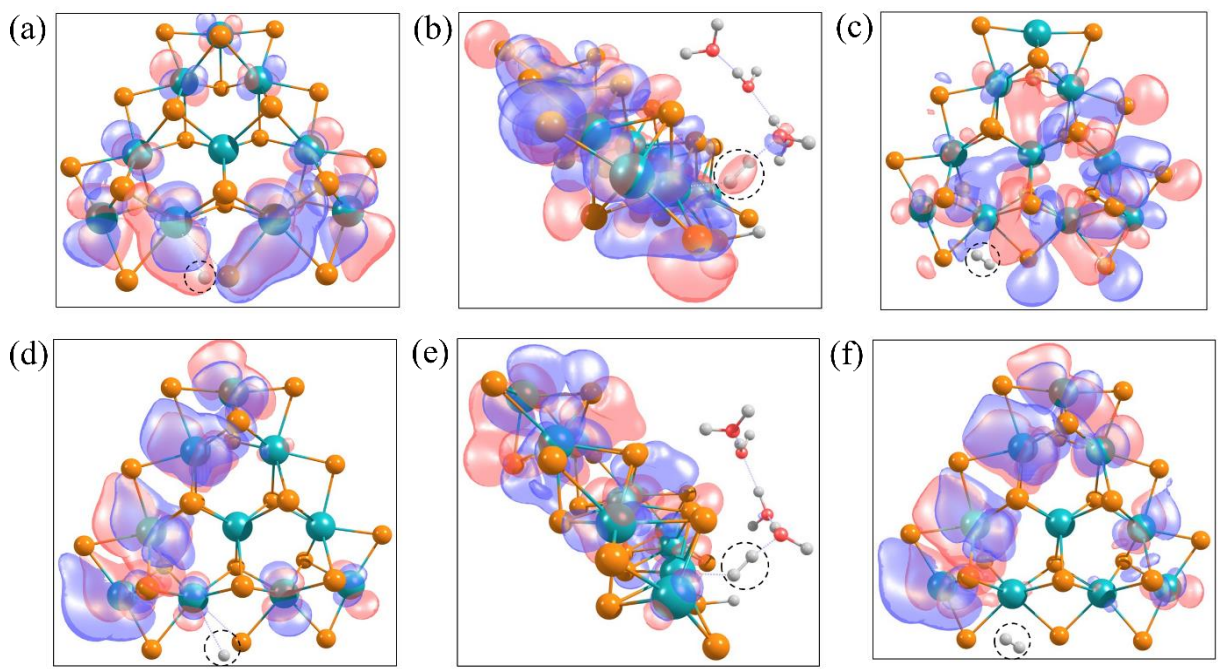

**Figure 13.** (a) HOMO of TS1 during the H$^\bullet$-migration, (b) HOMO of TS2 during the Heyrovsky reaction step for H$_2$ evolution, (c) HOMO of TS3 during the Tafel reaction step for H$_2$ evolution, (d) LUMO of TS1 during the H$^\bullet$-migration, (e) LUMO of TS2 during the Heyrovsky reaction step for H$_2$ evolution and (f) LUMO of TS3 during the Tafel reaction step of the H$_2$ evolution reaction taken place on the surfaces of the pristine 2D WSe$_2$ TMD material are shown here.

The Tafel slope (***b***) shows how efficiently an electrode can produce current in response to change in applied potential. So, if the Tafel slope (mV/decade) is lower means less overpotential is required to get high current assuming that the reaction rate of the catalyst does not limit with the electron transferred from the support to the catalyst. Theoretically, the Tafel slope (represented by "***b***") is given as $\boldsymbol{b} = 2.303 \left(\frac{RT}{nF}\right)$, where R= universal gas constant, T = temperature in the absolute scale, F=faraday constant (96485 C mol$^{-1}$) and n is the number of electrons transferred to the system during the HER process. In other words, it is an equation in electrochemical kinetics relating the rate of an electrochemical reaction to the overpotential. This Tafel slope provides the information regarding the rate-determining step, kinetics, energy required to obtain the required activity, etc., of the electrocatalysts. Tafel slope is an inverse measure of how strongly the reaction rate responds to changes in potential. It is used to evaluate the rate determining steps during the HER generally assume extreme coverage of the adsorbed species. The preset DFT study has found that the Tafel slop is about 29.583 mV dec$^{-1}$ as T = 298.15 K and n = 2 (for two electrons transfer) during the evolution of one H$_2$ molecule considering the non-periodic finite molecular cluster model calculations of the 2D WSe$_2$ TMD.

**Table 5. Reaction barriers and TOF for various 2D TMDs.**

| Catalysts | Heyrovsky TS2 barrier | | Tafel TS3 barrier | | TOF for Heyrovsky TS2 (solvent phase, in sec$^{-1}$) | References |
|---|---|---|---|---|---|---|
| | Gas phase (kcal/mol) | Solvent phase (kcal/mol) | Gas phase (kcal/mol) | Solvent phase (kcal/mol) | | |
| **MoS$_2$** | 16.0 | 23.8 | - | - | 2.1x10$^{-5}$ | 9 |
| **W$_{0.4}$Mo$_{0.6}$S$_2$** | 11.5 | 13.3 | - | - | 1.1x10$^3$ | 9 |
| **WS$_2$** | 14.5 | 21.3 | - | - | 1.5x10$^{-3}$ | 9 |
| **WSe$_2$** | 6.24 | 8.41 | 4.56 | 6.61 | 4.2x10$^6$ | Present work |

The activation barriers of both the TS2 (Heyrovsky reaction step) and TS3 (Tafel reaction step) during the H$_2$ formation in HER process on the surfaces of various TMDs (like 2D MoS$_2$, WS$_2$, W$_{0.4}$Mo$_{0.6}$S$_2$ and pristine 2D WSe$_2$) are shown in Table 5 and the values of TOFs are shown in Table 5 for assessment. A comparison has been drawn between them about their HER performances considering these activation barriers with the value of TOFs. Table 4 and 5 shows that the pristine 2D WSe$_2$ material has the lowest activation barriers during the H$^{\bullet}$-migration, H$_2$ evolution in the Heyrovsky and Tafel reaction steps with a higher value of TOF compared with other reported 2D TMDs as catalysts. These lower activation energies, higher TOF and lower Tafel slope of the pristine 2D monolayer WSe$_2$ give an insight into the superior electrocatalytic activity for the efficient HER compared to other reported 2D TMDs.

## 3. CONCLUSIONS

In summary, the equilibrium 2D monolayer structure, geometry, and electronic properties (such as band structures, band gap, Fermi energy level and total DOS) with electrocatalytic performance of the pristine 2D monolayer WSe$_2$ TMD material have been studied by employing first principles-based hybrid DFT method. The electronic property calculations found that the pristine 2D monolayer WSe$_2$ is a direct band gap semiconductor with a band gap about 2.39 eV at the K point in the $\Gamma$-$M$-$K$-$\Gamma$ high symmetric direction of the irreducible Brillouin zone. A non-periodic finite molecular cluster model system W$_{10}$Se$_{21}$ has been developed to explore the most efficient HER mechanism on the active surfaces of the 2D monolayer WSe$_2$ material by performing both the possible Volmer-Heyrovsky and Volmer-Tafel reaction pathways at the W-edges ($10\bar{1}0$). The catalytic performance of the pristine 2D monolayer WSe$_2$ TMD has been explored by computing the reaction barriers corresponding to the Gibbs free energy change during the H$^{\bullet}$-migration and H$_2$ evolution on the active surfaces. The present study has found that the reaction barriers of the TS1 and TS2 are about 6.11-8.41 kcal/mol in the Volmer-Heyrovsky mechanism during the HER process. The reaction barrier corresponding to the TS3 in the Tafel reaction step of the Volmer-Tafel reaction pathway is about 4.56-6.61 kcal/mol in the gas and solvent phases which is lower than the others. These computed single digit values of the activation reaction barriers in both the Volmer-Heyrovsky and Volmer-Tafel mechanisms of the HER process on the surfaces of pristine 2D monolayer WSe$_2$ TMD is lower than other previously reported 2D TMDs. This confirms the excellent catalytic activity of the pristine 2D monolayer WSe$_2$ compared to the other TMDs such as pristine 2D WS$_2$, MoS$_2$ and their W$_x$Mo$_{1-x}$S$_2$ alloys. HOMO-LUMO calculations have been performed at the equilibrium structures of the TSs (TS1, TS2 and TS3; i.e. the TSs formed in the subject reactions; H$^\bullet$-migration and the H$_2$ evolution either in the Heyrovsky reaction step or in the Tafel step) during the HER process. The overlap of the *s*-orbitals of the hydrogen atom attached with the W in the pristine 2D monolayer WSe$_2$ TMD and the water cluster (3H$_2$O + H$_3$O$^+$) seemed in the HOMO-LUMO Heyrovsky's transition state TS2 during H$_2$ formation has been found in the present study, and this better overlap of the atomic orbitals during the H$_2$ creation in the Heyrovsky's TS2 reduces the reaction barrier. This stabilization of the reaction limiting step in both the gas and solvent phases is a key for reducing both the H$^\bullet$-migration and Heyrovsky's reaction energy barriers, which results in a better electrocatalytic performance for HER compared to the ordinary TMDs. This illustrates why the 2D monolayer WSe$_2$ material has superior HER catalytic activity. The value of TOF was found about $4.24 \times 10^6$ sec$^{-1}$ during the Heyrovsky reaction step, and $8.86 \times 10^7$ sec$^{-1}$ during the Tafel reaction step at the active W-edges. These higher values of TOF confirm the efficient amount of H$_2$ evolution per active site of the pristine 2D monolayer WSe$_2$ catalysts per unit time. The theoretical value of the Tafel slope was found about 29.58 mV.dec$^{-1}$ during the HER process computed by the DFT method. Therefore, the low values of reaction barriers during adsorbed hydrogen migration and molecular hydrogen formation, Tafel slope and the ultra-high value of TOF altogether confirm that the pristine 2D monolayer WSe$_2$ has an excellent electrocatalytic activity for HER. It can be used as a promising and efficient noble metal-free HER catalyst for the efficient production of H$_2$.

## Supporting Information

The Supporting Information is available free of charge at https://pubs.acs.org/

Supporting information S1 (overview of the computational details), S2 (energy barriers and TOF comparison table), S3(optimized structure of symmetric 2D WSe$_2$ (.cif format)), and S4 (optimized geometries of non-periodic relevant structures in gas phase) are available**.**

## Author Contributions:

Dr Pakhira developed the complete idea of this current research work, and he computationally investigated the equilibrium structures and electronic properties of the pristine 2D monolayer WSe$_2$ material. Dr. Pakhira explored the whole reaction pathways; transitions states and reactions barriers and he explained the HER mechanism by the DFT calculations. Quantum calculations and theoretical models were designed and performed by Dr Pakhira and Dr Pakhira wrote the whole manuscript and prepared all the tables and figures in the manuscript. Dr. Pakhira and Mr. Vikash Kumar interpreted and analyzed the computed results and Dr Pakhira supervised the project work.

## AUTHOR INFORMATION

**Corresponding Author**
**Dr. Srimanta Pakhira** − *Department of Physics, Indian Institute of Technology Indore (IIT Indore), Simrol, Khandwa Road, Indore, Madhya Pradesh 453552, India.*

*Department of Metallurgical Engineering and Materials Science (MEMS), Indian Institute of Technology Indore (IITI), Simrol, Khandwa Road, Indore, Madhya Pradesh 453552, India.*

*Centre for Advanced Electronics (CAE), Indian Institute of Technology Indore, Simrol, Khandwa Road, Indore, Madhya Pradesh 453552, India.*

ORCID: orcid.org/0000-0002-2488-300X.
Email: spakhira@iiti.ac.in or spakhirafsu@gmail.com

**Author**

**Mr. Vikash Kumar** − *Department of Physics, Indian Institute of Technology Indore (IIT Indore), Simrol, Khandwa Road, Indore, Madhya Pradesh 453552, India.*

ORCID: orcid.org/0000-0002-8811-0583

## Acknowledgment:

Authors acknowledges the Science and Engineering Research Board-Department of Science and Technology (SERB-DST), Govt. of India for funding and supporting this research work by under Grant No. ECR/2018/000255 and CRG/2021/000572. Dr. Srimanta Pakhira is grateful for the financial support from the SERB-DST, Govt. of India under the scheme number ECR/2018/000255. Dr. Pakhira thanks to the SERB for providing highly prestigious Ramanujan Faculty Fellowship under the scheme number SB/S2/RJN-067/2017. Dr. Pakhira thanks to the SERB for providing highly prestigious Core Research Grant (CRG), SERB-DST, Govt. of India under the scheme number CRG/2021/000572. Mr. Vikash Kumar thanks the Indian Institute of Technology Indore (IIT Indore) and UGC, Govt. of India for availing his doctoral fellowship UGC Ref. No: 1403/ (CSIR-UGC NET JUNE 2019). The author would like to acknowledge the SERB-DST for providing the computing cluster and programs, and IIT Indore for providing the basic infrastructure to conduct this research work. We gratefully acknowledge the support and the resources provided by PARAM Brahma Facility under the National Supercomputing Mission (NSM), Government of India at the Indian Institute of Science Education and Research, Pune, India.

**Conflicts of Interest:**

The authors have no additional conflicts of interest.